\documentclass[times,amssymb,usenatbib]{mn2e}
\usepackage{psfig}

\newcommand{\be}{\begin{equation}}
\newcommand{\en}{\end{equation}}

\def\zabs{$z_{\rm abs}$}
\def\zem{$z_{\rm em}$~}

\def\kms{km~s$^{-1}$}
\def\h2{H$_2$}
\title[DLAs towards J1337+3152]{Detection of 
21-cm, H$_2$ and Deuterium absorption at $z>3$ 
along the line-of-sight to J1337+3152\thanks{Based on observations carried out at the European Southern Observatory (ESO), under programmes 082.A-0544 and
083.A-0454 (PI: C. Ledoux) with the UVES echelle spectrograph installed at the ESO Very Large Telescope (VLT), unit Kueyen, on Mount Paranal in Chile}}
\author[Srianand et al.]{R. Srianand$^{1}$\thanks{E-mail: anand@iucaa.ernet.in},
N. Gupta$^2$, P. Petitjean$^3$, P. Noterdaeme$^1$ \& C. Ledoux$^4$
\\
$^{1}$ IUCAA, Postbag 4,  Ganeshkhind, Pune 411 007, India \\
$^{2}$ Australia Telescope National Facility, CSIRO, Epping, NSW 1710, Australia \\
$^{3}$ Universit\'e Paris 6, UMR 7095, Institut d'Astrophysique de Paris-CNRS, 98bis Boulevard Arago, 75014 Paris, France \\
$^{4}$ European Southern Observatory, Alonso de C\'ordova 3107, Casilla 19001, Vitacura, Santiago 19, Chile \\
}

\begin{document}

\date{Accepted. Received; in original form }

\pagerange{--} \pubyear{2010}

\maketitle

\label{firstpage}

\begin{abstract}
We report the detection of 21-cm and molecular hydrogen absorption lines in
the same damped Lyman-$\alpha$ system (with log 
$N$(H~{\sc i})=21.36$\pm$0.10) at 
\zabs=3.17447 towards SDSS~J133724.69+315254.55 (\zem$\sim$3.174). We estimate 
the spin temperature of the gas to be, $T_{\rm S}=600^{+222}_{-159}$~K, 
intermediate between the expected values for cold and warm neutral media. 
This suggests that the H~{\sc i} absorption originates from a mixture of 
different phases. The total  molecular fraction is low, 
$f_{\rm H_2}$=10$^{-7}$, and \h2 rotational level populations 
are not in equilibrium.
The average abundance
of the  $\alpha$-elements 
is, [S/H]=$-1.45\pm0.22$.
Nitrogen and iron are 
found underabundant with respect to $\alpha$-elements by $\sim$1.0~dex and 
$\sim$0.5~dex respectively.
Using photoionization models we  conclude that the gas,
of mean density, ${n_H}\sim$2~cm$^{-3}$, is located more than 270
kpc away from the QSO.
While the position of  21-cm absorption line coincides with the
\h2 velocity profile, its centroid is shifted by
$\sim$2.7$\pm$1.0~\kms~ with respect to the redshift measured from
the \h2 lines.
However, the position of the strongest
metal absorption component matches the position of the
21-cm absorption line within 0.5~\kms. From this, we constrain the variation
of the combination of fundamental constants $x = \alpha^2 G_{\rm p}/\mu$,
$\Delta x/x = -(1.7\pm1.7)\times 10^{-6}$. 
This system is unique as we can at the same time 
have an independent constrain on $\mu$ using \h2 lines.
However, as the \h2 column density is low, only Werner 
band absorption lines are seen and, unfortunately, 
the range of sensitivity coefficients is too narrow to provide a stringent constraint: 
$\Delta\mu/\mu \le 4.0 \times 10^{-4}$.
The Ultraviolet and Visual Echelle Spectrograph (UVES)
spectrum reveals another DLA at \zabs = 3.16768 with
log $N$(H~{\sc i}) = 20.41$\pm$0.15 and low metallicity, [Si/H]~=~$-2.68\pm0.11$, in which
[O/C]~$\sim$~$0.18\pm0.18$ and [O/Si]~$\sim$~0. 
This shows that even in the very early stages of chemical evolution, 
the carbon or silicon to oxygen ratios can be close to solar. 
Using Voigt profile fitting we derive
log($N$(D~{\sc i})/$N$(H~{\sc i}))~=~$-(4.93\pm0.15)$ in this system.
This is a factor of two smaller than the value expected from the best fitted
value of $\Omega_{\rm b}$ from the Wilkinson Microwave Anisotropy
Probe (WMAP) 5 year data. This confirms the presence
of astration of deuterium even at very low metallicity.

\end{abstract}


\section{Introduction}

Damped Lyman-$\alpha$ systems (DLAs) are the highest H~{\sc i} column
density  absorbers seen in QSO spectra, with 
$N$(H~{\sc i})$\ge 10^{20}$ cm$^{-2}$, over cosmological time-scales. These absorbers trace 
the bulk of neutral hydrogen at $2\le z\le 3$ 
\citep [for example see,][]{Noterdaeme09dla} and have long been identified as 
revealing the interstellar medium of the high-redshift precursors of 
present day galaxies \citep[for a review see,][]{wolfe05}.
The physical conditions in the diffuse interstellar gas are
influenced by a number of physical processes including
in-situ star-formation, cosmic ray interaction, photoelectric heating by dust 
as well as mechanical energy input from both impulsive disturbances like 
supernova explosions and steady injection of energy in the 
form of stellar winds.  
Therefore, studies of the physical conditions in DLAs 
provide a probe of the cosmological evolution of normal galaxies
selected solely based on H~{\sc i} cross-section. 

Our understanding of physical conditions in DLAs is primarily
derived from optical absorption-line studies, involving the detection of
low-ionization metal transitions and, in some cases, molecular absorption
[e.g. \citet{petitjean00}, \citet{ledoux03} and \citet{noterdaeme08}, for H$_2$, and 
\citet{srianand08} and \citet{noterdaeme09co} for CO].
The gas in which H$_2$ is detected, has typical
temperature and density of 150$\pm$80~K and
$n_{\rm H}$ = 10$-$200 cm$^{-3}$ respectively. 
These conditions correspond to those prevailing in a cold neutral medium (CNM) 
embedded in a radiation field of moderate intensity originating from local star formation activity \citep{Srianand05}.

Detecting 21-cm absorption in DLAs is one complementary way to probe the
physical conditions in the absorbing gas. If detected, the 21-cm optical 
depth, in conjunction with the observed $N$(H~{\sc i}) from optical 
data, gives the spin temperature of the H~{\sc i} gas \citep[see][]{kanekar03}. 
It is widely believed that the latter is a good tracer 
of the kinetic temperature. 
The width of the 21-cm line observed at very high spectral
resolution can also yield a direct measurement (or give a 
stringent upper limit on) of the kinetic temperature \citep[as,
for example, in the \zabs = 1.3603 system towards J~2340$-$0053,][]{gupta09}. 
This can be combined 
with measurements of C~{\sc i}, C~{\sc ii}, Si~{\sc ii} and/or O~{\sc i}
fine-structure lines
to derive the gas density \citep{Heinmuller06}.

By measuring the relative positions of absorption lines it is
possible to constrain the time variation of several 
dimensionless constants. Accurately measured wavelengths of \h2 absorption 
lines
allow one to constrain
the variation of the proton-to-electron mass
ratio \citep{Varshalovich93}.
The position of the 21-cm absorption line can be compared to those of 
the corresponding metal line components to probe the variations of a 
combination, $x=\alpha^2 G_{\rm p}/\mu$, of the fine structure constant
($\alpha$), the proton-to-electron mass ratio ($\mu$) and the proton
 gyromagnetic factor \citep[$G_{\rm p}$,][]{Tubbs80}.
Simultaneous detection of H$_2$, 21-cm and metal absorption line
in the same system can, in principle, allow one to lift part of the 
degeneracy between the  variations of these different constants.

\par
Despite obvious advantages, till recently,  there was 
only one system \citep[at $z_{\rm abs}$~=~1.7765 towards Q~1331+170,][]{Cui05}
known to produce 
both H$_2$ and 21-cm absorption lines. 
Due to the low redshift of the system, however, it is not possible to cover the 
H$_2$ Lyman and Werner band absorption with high resolution spectroscopy at high enough
signal-to-noise ratio. This prevents one from
performing a detailed analysis of the cloud.
In the course of our on-going GBT/GMRT survey for 21-cm absorption in high-$z$ DLAs selected from the 
SDSS we discovered a DLA system, exhibiting both 21-cm and H$_2$ absorption
lines, at
\zabs~=~3.17447 towards J133724.69+315254.55 (called J~1337+3152 hereafter).
  This system is unique
as both 21-cm absorption and \h2 are extremely rare in DLAs at $z\ge$3.
In addition, 
there is another DLA at \zabs~=~3.16768 along the same line-of-sight
(at $\sim$500~km~s$^{-1}$ away from the first DLA) with very low
metallicity.
In this paper, we present a detailed analysis of both DLAs.

\section{Details of observations}
\begin{figure}
\centerline{{
\psfig{figure=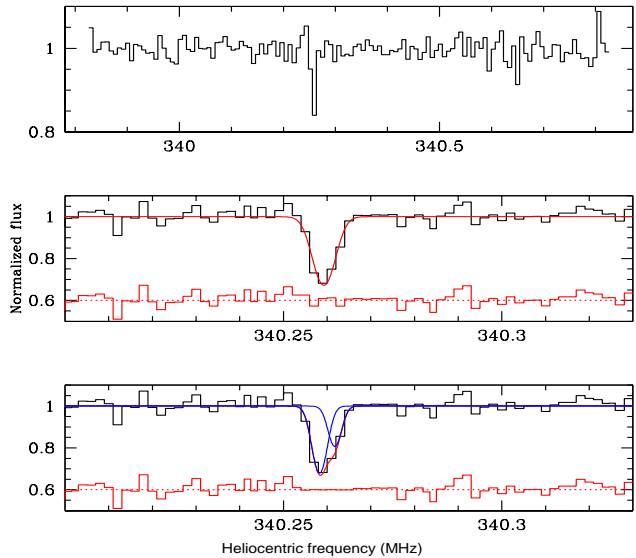,height=8.0cm,width=9.0cm,angle=0}
}}
\caption[]{Normalised GMRT spectra showing the detection of $z=3.174480$ 
21-cm absorption toward J~1337+3152. 
{\sl Top panel}: low resolution 
spectrum (6.88 km/s/channel) observed on January 13, 2009. 
{\sl Middle and bottom panels}: high resolution spectrum
(1.72 km/s/channel) together with, respectively, the best single or two gaussian 
component fit models. Residuals (shifted by 0.6 along the y-axis) are also 
shown.  
}
\label{21cmabs}
\end{figure}

The quasar J1337+3158 was observed with the GMRT on January 13, 2009 to search for 21-cm 
absorption in the DLA at \zabs=3.17448. 
Observations were performed in a bandwidth of 1\,MHz split into 128 channels.  
We observed the nearby standard flux density calibrator 3C\,286 every 45\,minutes for 
the amplitude, phase and bandpass calibration. A total of 6.2\,hrs of data 
were acquired on-source in both the circular polarization channels, RR and LL.
Data were reduced using the NRAO Astronomical Image Processing System (AIPS) 
following standard procedures as described in \citet{Gupta06}.  
Special care was taken to exclude the baselines and timestamps affected 
by radio frequency interferences (RFI).  
An absorption feature confined in a single spectral resolution channel  
(6.88\,km\,s$^{-1}$) is detected consistently in both polarisations
(see top panel of Fig.~\ref{21cmabs}).  In the Stokes I spectrum, the 
absorption corresponds to an integrated 21-cm optical depth, 
$\int\tau dv$~=~1.55$\pm$0.19 \kms. 
The quasar was 
reobserved on March 17, 2009 during 7.8\,hrs (on-source time) 
at a higher frequency resolution to confirm the 21-cm absorption. A bandwidth 
of 0.25\,MHz split into 128 channels, i.e. a spectral resolution of 1.72\,km\,s$^{-1}$,
was used. With respect to the first observing run, this second epoch observation 
corresponds to an overall shift in the observing frequency of 
18.2\,km\,s$^{-1}$ due to the heliocentric motion of the Earth. The absorption 
profile was observed to be consistent with this velocity shift.  
We used AIPS task CVEL to correct the observed data for the Earth's motion and 
rotation. The higher resolution Stokes I spectrum (with rms of 
2.0\,mJy\,beam$^{-1}$\,channel$^{-1}$) from the second epoch observation 
together with the fits to the H~{\sc i} absorption
are shown in the middle and bottom panels of Fig.~\ref{21cmabs}.

We also observed J~1337+3152  
with the Ultraviolet and Visual Echelle  spectrograph
(UVES; \citet{Dekker00}) in visitor mode on March 27 and April 27, 2009.
{We used two settings with two different central wavelengths 
 in the blue (390nm and 437nm)
and  red (600nm and 770nm) arms to cover the wavelength range 3300$-$9590~{\AA} with a single
gap between 7590 and 7769~{\AA}}. The CCD pixels were binned $2\times 2$ 
and the slit width adjusted to $1\arcsec$, yielding a resolving power of 
$R=45000$ under median seeing conditions of $\sim 0\farcs 9$. 
The total exposure
time on source is 4~hrs.
{ The data were reduced using the UVES pipeline 
\citep{Ballester00}, which is available in the context of
the ESO MIDAS data reduction system.}
Wavelengths
were rebinned to the heliocentric rest frame and individual scientific
exposures co-added using a sliding window and weighting the signal by
the inverse of the variance in each pixel. The resultant combined spectrum
has a SNR in the range 10-20 per pixel. We fitted the quasar continuum 
using low order polynomials in line free continuum regions.
We analysed the spectrum using standard Voigt profile fitting procedures.
The atomic data used are from \citet{Morton03} unless otherwise 
specified. {In the following, we adopt the  solar photospheric abundances 
from \citet{Asplund09}, rest wavelengths and oscillator strengths
of \h2 transitions from  \citet{Ubachs07} and \citet{Abgrall94} respectively}.

\section{High $N$(H~{\sc i}) systems}
\begin{figure*}
\centerline{{
\psfig{figure=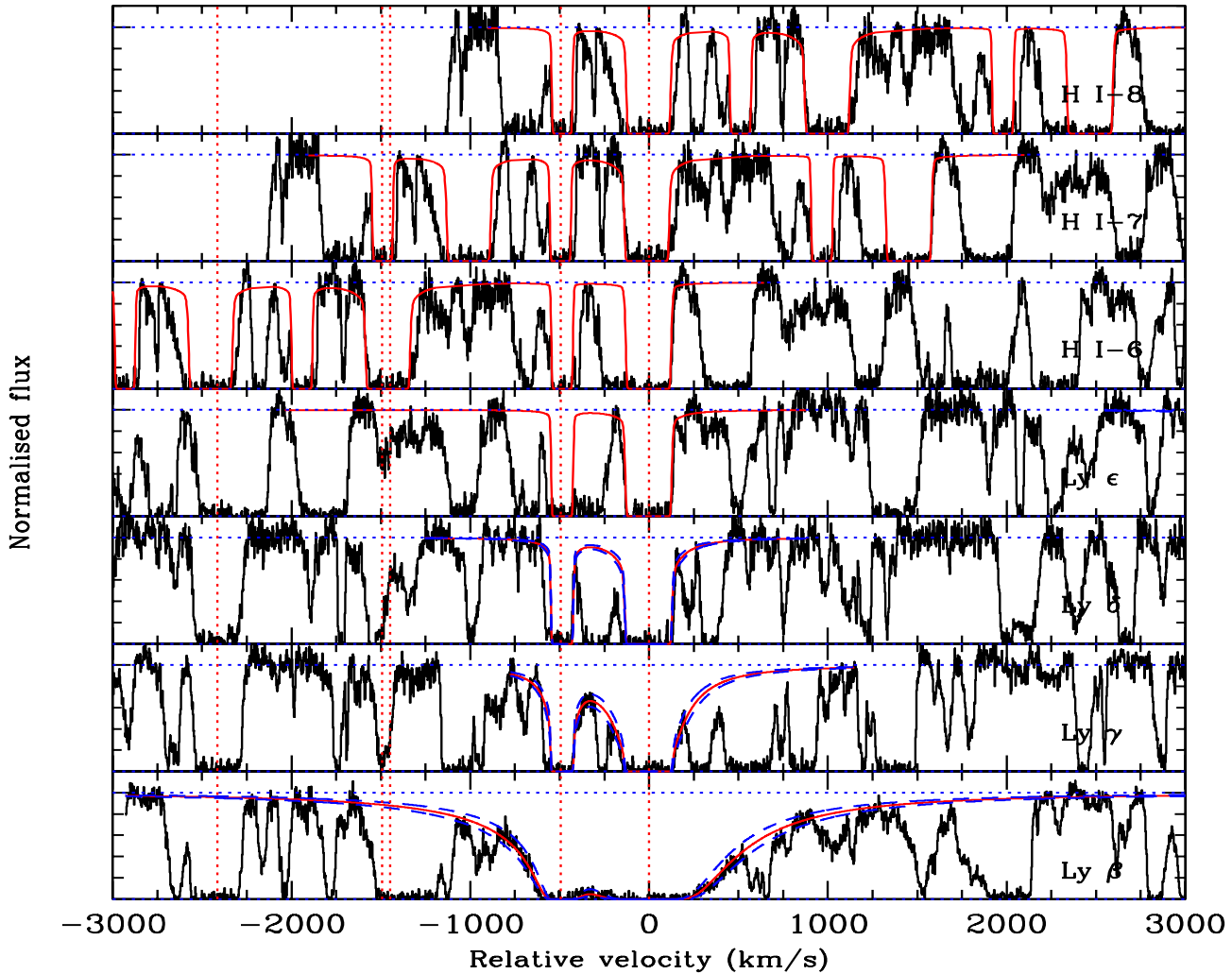,height=15.0cm,width=18.0cm,angle=0}
}}
\caption[]{Fit to the Lyman series lines of the $z_{\rm abs}$~=~3.17448 
($v=0$~\kms)  and 3.16768 ($v=-488$~\kms) DLA systems. There is a strong Lyman limit system at
$z_{\rm abs}$~=~3.1408 ($v=-2420$~\kms). Some of the Lyman series absorption lines
of the $z_{\rm abs}$~=~3.17448 and 3.1408 system happen to be blended. The
vertical dashed lines at $v\sim -1500$ \kms mark the locations of a 
C~{\sc iv} system along this line of sight (see also Fig.~\ref{highion}).
}
\label{lyman}
\end{figure*}

There are three high-$N$(H~{\sc i}) absorption systems within 2500~\kms~
along this line of sight at $z_{\rm abs}$~=~3.17448 ($v=0$~\kms), 
3.16768 ($v=-488$~\kms)
and 3.1408 ($v=-2420$~\kms). The corresponding Lyman series absorption lines
are shown on a velocity scale in Fig.~\ref{lyman}.
The Lyman-$\alpha$ profile is dominated by the \zabs~=~3.17448 system. As the red wing of this absorption 
is affected by the QSO Lyman-$\alpha$ emission, we do not use it to derive $N$(H~{\sc i})
but use instead Lyman-$\beta$ and other higher order lines.

The Lyman-$\beta$ line of the $z_{\rm abs}$~=~3.17448 and 3.16768 systems (referred to
as system-1 and system-2 from now on) are blended together.
Some of the higher order Lyman lines of system-1 are blended with lines from the sub-DLA at
\zabs = 3.1408 ($\sim -2420$ \kms).
The main constraint on log~$N$(H~{\sc i}) in system-1 
comes from the red wing of the Lyman-$\beta$ absorption and the core 
of the Lyman-$\gamma$ absorption. In the case of system-2, the blue wing
of Lyman-$\beta$ and the red wing of Lyman-$\gamma$ are used for the same
purpose. We obtain log~$N$(H~{\sc i})= 21.36$\pm$0.10 and 20.41$\pm$0.15 for
system-1 and system-2 respectively.
Note that errors  come mostly from continuum placement 
uncertainties. We discuss the fits to the H~{\sc i} lines for system-2 in greater detail below
(see Sect.~5).
In both cases, the unblended wings of the higher Lyman series lines are used
to constrain the velocity dispersion. 

In addition, based on the absence of damping wings in Lyman-$\beta$ and the blue wing
of the Lyman-$\alpha$ absorption lines we derive an upper limit of 
$10^{19}$~cm$^{-2}$ for $N$(H~{\sc i}) at \zabs = 3.1408 ($\Delta v$~$\sim~-2420$~\kms).
We cannot perform a more precise measurement as all the available 
transitions are saturated, hiding the component structure.
We therefore do not discuss this system any further.

\section{Analysis of System-1  (\zabs = 3.17448)}

Weak \h2 and strong 21-cm absorptions are seen associated with this system
in the UVES and GMRT spectra, respectively.

\subsection{21-cm absorption}

The GMRT spectrum covers only the expected 21-cm range for system-1. 
In the higher resolution GMRT spectrum the absorption is spread over 
7-8 channels and can be adequately fitted with a single Gaussian 
component of peak optical 
depth $\tau$~=~0.40$\pm$0.02 located at 340.2593$\pm$0.0002\,MHz (or
\zabs~=~$3.174480\pm0.000002$) and FWHM~=~5.08$\pm$0.34\,km\,s$^{-1}$ (see middle 
panel in Fig.~\ref{21cmabs}).

The gaussian fit was carried out using an IDL 
code based on MPFIT, which performs $\chi^2$-minimisation 
by Levenberg-Marquardt techniques \citep{Markwardt09}. We use the rms flux
measured in the line free channels as typical error in each channel.
The best fit has a reduced $\chi^2$ = 1.02.
The FWHM of the fitted Gaussian component corresponds to a kinetic temperature 
of $T_{\rm K}\le564\pm$76\,K.

Motivated by the fact that the centroid of the \h2 absorption does not coincide
with the centroid of the 21-cm absorption (see below), we performed a fit with two components.
The result of this fit is shown in the bottom panel of Fig~\ref{21cmabs}.
The best fitted two components have
$z_{\rm abs}$~=~3.174492$\pm$0.000007 
and 3.174451$\pm$0.000012 with FWHM~=~3.4$\pm$0.9 and 3.1$\pm$1.6 \kms~ 
respectively. These values translate to upper limits on kinetic temperatures 
of 252 and 210~K
respectively. The reduced $\chi^2$ in this case is 0.87. This, as well as
the residuals seen in Fig.~\ref{21cmabs}, suggest that the two component 
model is probably over-fitting the data. 

The H~{\sc i} column density of an optically thin cloud covering a
fraction $f_{\rm c}$ of a background radio source is related to the 
21-cm optical depth $\tau(v)$ in a velocity interval ($v$, $v+$d$v$) 
and to the spin temperature ($T_{\rm s}$) by 
\begin{equation}
N{\rm(H~I)}=1.835\times10^{18}~{T_{\rm s}\over f_{\rm c}}\int~\tau(v)~{\rm d}v~{\rm cm^{-2}}.
\label{eq1}
\end{equation}
The total optical depth is 
$\int\tau(v) dv$ = 2.08$\pm$0.17 \kms. Thus the total 
H~{\sc i} column density is, $N$(H~{\sc i})~=~(3.82$\pm$0.31)$\times10^{18}$($T_{\rm s}/f_{\rm c}$)\,cm$^{-2}$.  
From our optical spectrum we measure log~$N$(H~{\sc i})~=~21.36$\pm$0.10.
For $f_{\rm c} = 1$, this yields $T_{\rm s} = 600^{+222}_{-159}$ K. Interestingly this
is very close to the upper limit we get from the width of the 
21-cm absorption line for the single gaussian fit.

There is no milli-arcsecond scale resolution radio map available for the J~1337+3152 field.   
The radio source is compact in our GMRT images at 340\,MHz with a resolution of 
10$^{\prime\prime}$$\times 8^{\prime\prime}$ and has a flux density of 69.3\,mJy.
In the VLA FIRST 1.4\,GHz image (of resolution 5$^{\prime\prime}\times5^{\prime\prime}$) the
radio source is represented by a single component with flux density 83\,mJy and a 
deconvolved size of 0.75$^{\prime\prime}\times0.12^{\prime\prime}$.
It is compact as well in our GMRT 610-MHz image with similar resolution 
with a flux density of $\sim$102\,mJy.  
The image of highest resolution for this source is from the JVAS/CLASS survey. In this 
VLA A-array 8\,GHz image of resolution 0.26$^{\prime\prime}\times0.22^{\prime\prime}$
the source is still compact with a flux density of 25.8\,mJy.  
In the GB6 catalog at 5\,GHz this radio source has a flux density of 29\,mJy, very similar 
to the CLASS/JVAS 8\,GHz image.
Therefore, from  the arcsecond and sub-arcsecond scale images available, the source appears to be core 
dominated with an indication of the low-frequency turnover. 
We have no reason to believe the covering factor is not close to unity. 

The spin temperature measured here is in between the expected
kinetic temperatures of  cold ($\le 300$ K) and warm (5000$-$8000 K)
neutral media \citep{Wolfire03}. Thus, what we measure
can be interpreted as the harmonic mean temperature of the multiphase medium along
the line of sight. Interestingly the spin temperature we measure here is consistent
with the mean spin temperature measured in very low redshift QSO-galaxy
pairs \citep{Carilli92}.

\subsection{Molecular Hydrogen}
\begin{figure}
\centerline{{
\psfig{figure=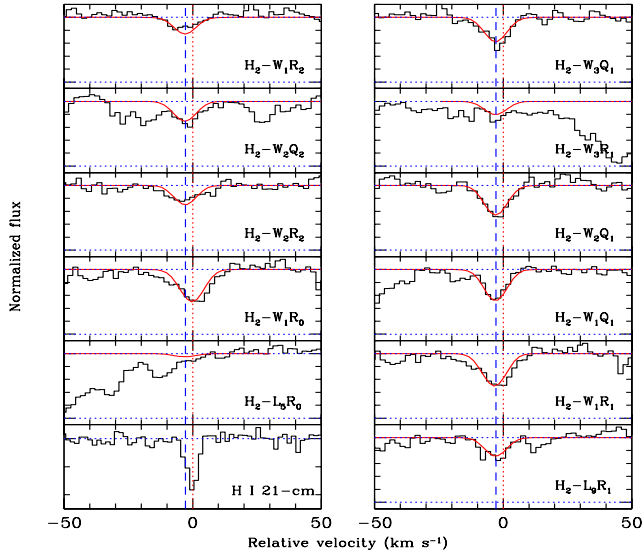,height=8.0cm,width=9.0cm,angle=0}
}}
\caption[]{Voigt profile fits to H$_2$ Lyman and Werner band absorption lines.
The vertical dotted and dashed lines mark the locations of the 
21-cm and H$_2$ absorption lines, respectively. The apparent shift 
between the two lines is  $\sim$2.7~\kms.{ The zero of the velocity
scale is defined with respect to the redshift of the \h2 lines, \zabs = 3.174441.}
}
\label{figh2}
\end{figure}

\begin{table}
\caption{Results of single component fits to H$_2$ lines in system-1}
\begin{tabular}{lccc}
\hline
Level & $z$ & $b$ & log $N$(H$_2$,J) \\
      &     & (\kms) & (cm$^{-2}$) \\
\hline
\hline
J=0 &3.174441(5) &4.6$\pm$0.4& 12.88$\pm$0.16\\
J=1 & ....                   &  ....   & 13.85$\pm$0.04\\
J=2 & ....                   &  ....   & 13.65$\pm$0.04\\
J=3 & ....                   &  ....   & $\le$ 13.47 \\
\hline
\end{tabular}
\label{tabh2}
\end{table}

Weak Lyman and Werner band H$_2$ absorption lines from different rotational levels
of the vibrational ground state are detected in system-1. 
The spectrum at the positions of different H$_2$ absorption lines
together with single component Voigt profile fits to these lines are shown in 
Fig.~\ref{figh2}. All the lines except H$_2$W$_3$R$_1$ and H$_2$L$_5$R$_0$ were 
used during the fits for which $z$ and $b$ were assumed to be the same for all $J$ levels. 
The reduced $\chi^2$ for the best fit is 1.4. Corresponding parameters are summarised 
in Table~\ref{tabh2}.

For J~=~0, most of the lines are blended in the Lyman-$\alpha$ forest and the quoted 
column density is constrained by the strength of the H$_2$W$_1$R$_0$ absorption.
The large error is due to the fact that this line is blended with H$_2$W$_1$R$_1$.
Similarly, in the case of J~=~3, most of the strongest transitions are blended with other lines
in the Lyman-$\alpha$ forest.  We use the expected position of the
H$_2$W$_3$R$_3$ line to derive an upper limit on $N$(H$_2$,J=3). 

From the total H$_2$ column density and the above measured
$N$(H~{\sc i}) we derive the molecular fraction in the cloud: 
$f_{\rm H_2} \simeq 2 N({\rm H_2})/N$({\rm H}~{\sc i}) = $1.0 \times 10^{-7}$
which is the lowest value measured in a DLA 
(see Table~1 of \citet{noterdaeme08}). The low $f_{\rm H_2}$
could be the result of either or both low dust content and excess UV flux 
from the QSO. 

The J~=~1 excitation temperature 
is usually used as an estimator of the kinetic temperature. However, it is known that this 
estimator is robust only when the total \h2 column density is in excess of 
10$^{16}$~cm$^{-2}$ \citep{Roy06}, which is far from being the case here. 
Indeed, the column density ratio $N$(H$_2$,J=1)/$N$(H$_2$,J=0) 
measured here is higher than, or close to 9, the maximum value allowed by the Boltzmann 
distribution. Note that this has already been observed in the low-$N$(H$_2$) 
\zabs~=~3.0248 system toward Q~0347$-$383 (see \citet{ledoux03}, \citet{Srianand05}).

From Fig.~\ref{figh2} it is apparent that, while the 21-cm absorption
is well within the velocity range of the \h2 absorption, the centroid of
the \h2 absorption lines is shifted by 2.7$\pm$1.0 \kms with respect
to the centroid of the 21-cm absorption line for the single component fit.
If we use the two component gaussian fit to the 21-cm line then the 
redshift of the weak 21-cm absorption 
component is consistent with that of the \h2 absorptions within measurement
uncertainties. The basic observation remains, however, that
the optical depth ratio between \h2 and 21-cm absorption 
varies inside the absorbing cloud.

\subsection{Heavy element abundances and dust content}
 
\begin{table}
\caption{Results of Voigt profile fits to the metal lines in system-1 at \zabs = 3.17447}
\begin{tabular}{l c c c c}
\hline
Species & $z$ & log~$N$   & $b$     & [X/S]$^a$\\
        &     & (cm$^{-2}$) & (\kms)  &      \\
\hline
\hline
N~{\sc i}   &  3.174025  & 13.78$\pm$0.03 & 11.3$\pm$ 0.6 & $-$0.95$\pm$0.05\\
Si~{\sc ii} & ....       & 14.60$\pm$0.09 & ....          &$+$0.26$\pm$0.10\\
S~{\sc ii}  & ....       & 13.98$\pm$0.04 & ....          &....\\
Fe~{\sc ii} & ....       & 13.90$\pm$0.02 & ....          &$-$0.38$\pm$0.09\\
\\
C~{\sc ii}$^*$  &  3.174387  & 13.34$\pm$0.04 & 8.3$\pm$ 0.4    &.... \\
C~{\sc i}   &....        & $\le$ 12.30    & .... \\
N~{\sc i}   &....        & 14.68$\pm$0.02 & ....            &$-$0.87$\pm$0.03\\
O~{\sc i}   &....        & 16.25$\pm$0.10 & ....            &$-$0.08$\pm$0.10\\
Mg~{\sc ii} &....        & 15.13$\pm$0.06 & ....            &$-$0.05$\pm$0.06\\
Si~{\sc ii} &....        & 15.17$\pm$0.04 & ....            &$+$0.00$\pm$0.04\\
S~{\sc ii}  &....        & 14.80$\pm$0.02 & ....            &....\\
Ar~{\sc i}  &....        & 13.62$\pm$0.06 & ....            &$-$0.38$\pm$0.10\\
Cr~{\sc ii} &....        & 13.00$\pm$0.17 & ....            &$-$0.29$\pm$0.17\\
Fe~{\sc ii} &....        & 14.53$\pm$0.03 & ....            &$-$0.57$\pm$0.08\\
Ni~{\sc ii} &....        & 13.41$\pm$0.04 & ....            &$-$0.44$\pm$0.04\\
Zn~{\sc ii} &....        & 12.17$\pm$0.25 & ....            &$-$0.10$\pm$0.25\\
\\
C~{\sc ii}$^*$ &  3.174473  & 13.27$\pm$0.07 & 2.5$\pm$ 0.6  &....\\ 
C~{\sc i}   &....        & $\le$ 12.30    & .... \\
N~{\sc i}   &....        & 14.12$\pm$0.15 &....           &$-$1.38$\pm$0.26\\
O~{\sc i}   &....        & 16.60$\pm$0.13 & ....          &$+$0.32$\pm$0.16\\
Mg~{\sc ii} &....        & 15.03$\pm$0.06 & ....          &$-$0.10$\pm$0.22\\
Si~{\sc ii} &....        & 15.11$\pm$0.11 & ....          &$+$0.00$\pm$0.24\\
S~{\sc ii}  &....        & 14.75$\pm$0.21 & ....          & ....\\
Ar~{\sc i}  &....        & 13.70$\pm$0.08 & ....          &$-$0.25$\pm$0.23\\
Cr~{\sc ii} &....        & 12.89$\pm$0.22 & ....          &$-$0.35$\pm$0.30\\
Fe~{\sc ii} &....        & 14.61$\pm$0.07 & ....          &$-$0.44$\pm$0.23\\
Ni~{\sc ii} &....        & 13.26$\pm$0.05 & ....          &$-$0.54$\pm$0.22\\
Zn~{\sc ii} &....        & $\le$12.30     & ....          &....\\
\hline
\end{tabular}
\begin{flushleft}
{$^a$ Solar photospheric abundances are from \citet{Asplund09}.}
\end{flushleft}
\label{voigt_met}
\end{table}

\begin{figure*}
\centerline{{
\psfig{figure=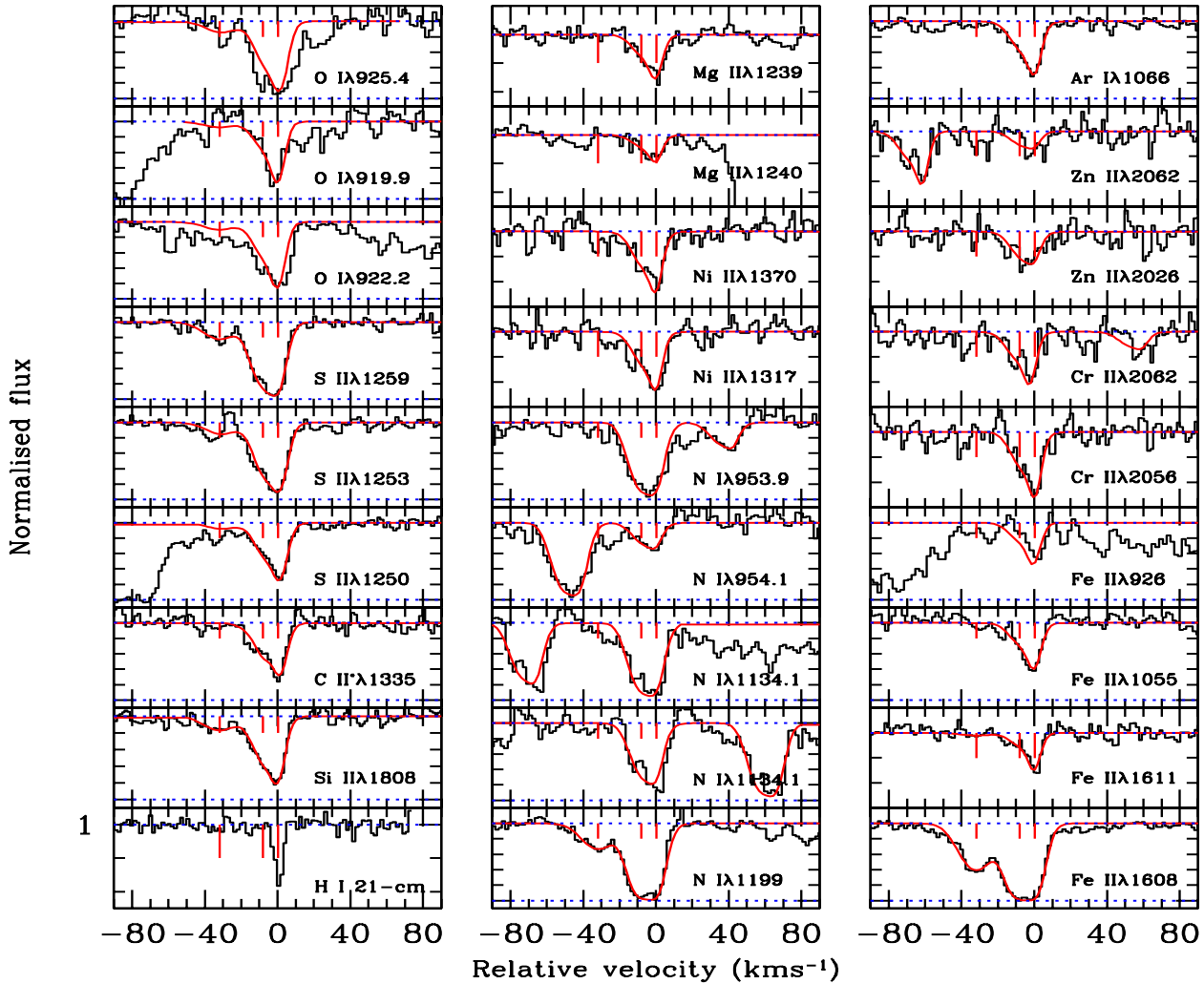,height=15.0cm,width=18.0cm,angle=0}
}}
\caption[]{Voigt profile fits to a selected set of metal absorption lines 
from system-1 at \zabs = 3.17447. Vertical tick marks indicate the locations 
of the three velocity components.
}
\label{metal_z1}
\end{figure*}

\begin{figure*}
\centerline{
\psfig{figure=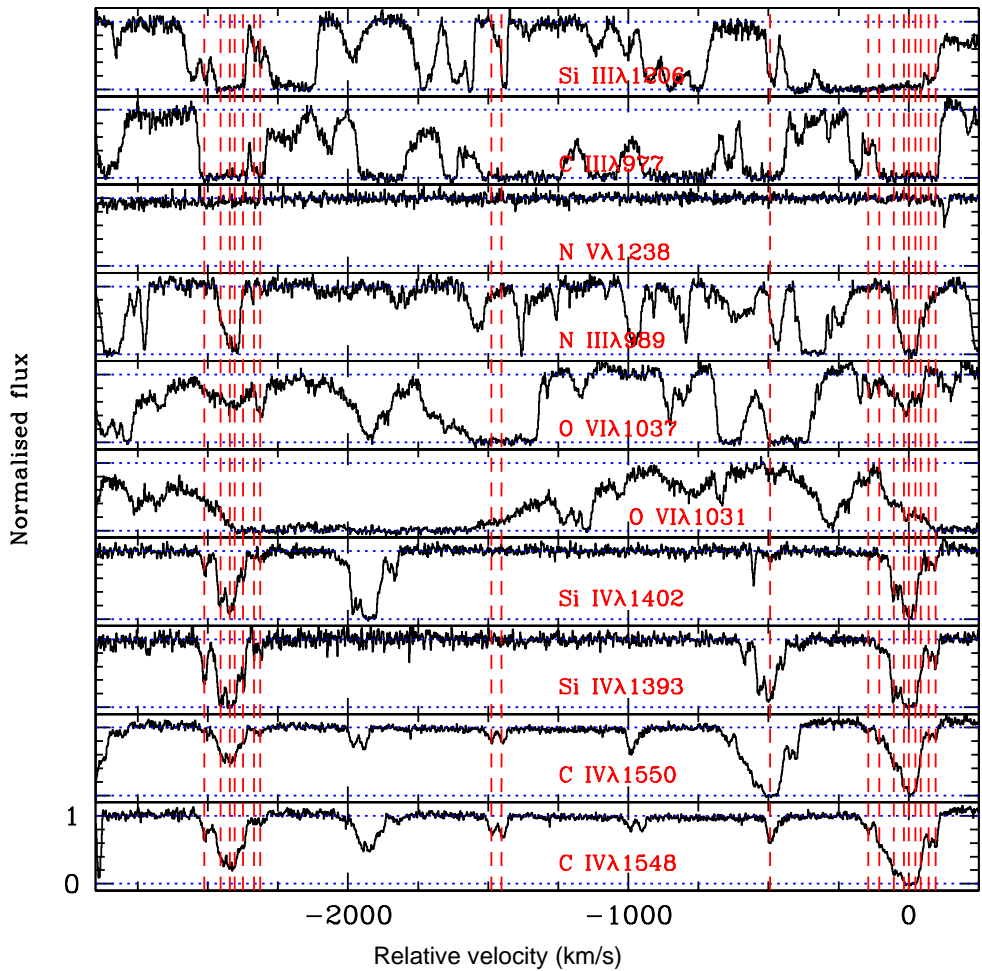,height=14.0cm,width=14.0cm,angle=0}}
\caption[]{Velocity plot of metal absorption lines of highly ionized 
species.
The velocity scale is defined with respect to the DLA at
\zabs~=~3.17447. Dashed vertical lines mark the locations of
C~{\sc iv} components. The system at $v\sim$$-$500~\kms 
corresponds to the DLA at \zabs~=~3.16768. The vertical 
dashed lines around $v\sim -2420$~\kms are the C~{\sc iv}
components of a LLS at \zabs = 3.1408. The C~{\sc iv}
components marked at $v\sim -1500$~\kms are from an intervening
high column density Lyman-$\alpha$ system. 
}
\label{highion}
\end{figure*}

\begin{table}
\caption{Average metallicities in system-1 at $z_{\rm abs}$~=~3.17447}
\begin{center}
\begin{tabular}{l c c}
\hline
\hline
Species & log~$N$   &  log~Z/Z$_\odot$$^a$ \\
        & (cm$^{-2}$) &                        \\
\hline
H~{\sc i}     & 21.36$\pm$0.10 & ....\\
C~{\sc i}     & $\le$ 12.60    & ....\\
C~{\sc ii$^*$}& 13.61$\pm$0.08 & ....\\
N~{\sc i}     & 14.83$\pm$0.06 & $-$2.44$\pm$0.12\\
O~{\sc i}     & 16.76$\pm$0.16 & $-$1.40$\pm$0.19\\
Mg~{\sc ii}   & 15.59$\pm$0.08 & $-$1.35$\pm$0.13\\
Si~{\sc ii}   & 15.50$\pm$0.15 & $-$1.42$\pm$0.18\\
S~{\sc ii}    & 15.11$\pm$0.20 & $-$1.45$\pm$0.22\\
Ar~{\sc i}    & 13.96$\pm$0.10 & $-$1.80$\pm$0.14\\
Cr~{\sc ii}   & 13.25$\pm$0.28 & $-$1.79$\pm$0.30\\
Fe~{\sc ii}   & 14.91$\pm$0.08 & $-$1.95$\pm$0.13\\
Ni~{\sc ii}   & 13.64$\pm$0.15 & $-$1.97$\pm$0.18\\
Zn~{\sc ii}   & 12.26$\pm$0.26 & $-$1.77$\pm$0.28\\
\hline

\end{tabular}

\begin{flushleft}
{$^a$ solar photospheric abundances are from \citet{Asplund09}.}
\end{flushleft}
\label{ave_abundance}
\end{center}
\end{table}

We performed simultaneous Voigt profile fits to metal absorption line  features 
from neutral and singly ionized species (see Fig.~\ref{metal_z1}). 
The best fit is obtained with three components at 
$z_{\rm abs}$~=~3.174025, 3.174387 and 3.174473. First 
two components have relative 
velocities compared to the third one of $\Delta v$~=~$-$32 and $-$6~km~s$^{-1}$ respectively.
Fit parameters are summarized in Table~\ref{voigt_met},
where the last column gives relative abundances with respect to Sulfur
in each component without applying any ionization correction. 

In  Fig.~\ref{metal_z1}, we also plot the 21-cm absorption profile
for comparison. It is clear from this figure that the strongest
and narrow metal line component at \zabs~=~3.174473(6) is well aligned 
with the 21-cm absorption. It is also clear from Table~\ref{voigt_met}
that there is no strong component to component variation in the 
depletion pattern among the three components which is consistent with what 
is observed in typical DLAs \citep{Rodriguez06}. 

Based on the robust column densities of Si~{\sc ii}, S~{\sc ii} and Fe~{\sc ii} we find
the main component at \zabs~=~3.174473 to contain 40 to 45\% of the total column density 
detected in this DLA. As there is no reason to believe the metallicity
in this component is less than in the other two components, it is most
likely that only 40 to 45\% of the total H~{\sc i} column density is associated
with this component. In that case we derive $T_{\rm s}\sim 270^{+100}_{-72}$~K 
for this component. This brings the spin temperature very close to what is 
expected in a cold neutral medium (CNM).
It is also interesting to note that there is no
indication of a distinct narrow metal component coinciding with the \h2 position
supporting the fact that the cloud is not strictly homogeneous and that
the \h2 and 21-cm narrow components are consequences of physical conditions in the cloud
changing on small scale.

Adding column densities from the three components, we derive the 
average metallicities in the DLA given in Table~\ref{ave_abundance}.
The $\alpha$ element metallicity is [Si/H]~$\sim$~[S/H]~=~$-1.45\pm0.22$. 
{The oxygen abundance measured from the unblended lines matches well 
(i.e [O/S] = 0.05$\pm$0.29) with that of other $\alpha$ elements although
the errors are large}.

The nitrogen abundance measurement is robust and 
we find [N/$\alpha$]~$\sim$~$-$1.0. This is consistent with 
what is seen in low metallicity 
DLAs \citep[see][]{Petitjean08,Pettini08a}
and very different of what is usually measured in gas associated with QSOs 
\citep[][]{Petitjean94, Hamann93}. 

From Table~\ref{ave_abundance}, it can be seen that Ar seems slightly
($\sim$0.4~dex) underabundant compared to other $\alpha$ 
elements. Such deficiency has been noted already in the Galactic ISM 
and in high redshift DLAs \citep{Vladilo03, Ledoux06b}. 
This has been attributed to the large photoionization cross-section of
Ar~{\sc i} above 15~eV \citep{Sofia98}. We can use this argument 
to constrain the ionization parameter in the absorbing gas (see below). Since 
the [Ar/Si] ratio is about the same in the individual components  
(see Table~\ref{voigt_met}), we can infer that the effect of ionization 
is nearly 
identical in the different components.

The Zn~{\sc ii} transitions are weak and the corresponding column
densities are difficult to measure. The [Zn/Cr] metallicity ratio 
is close to solar while Fe, Cr and Ni are slightly underabundant 
compared to S and Si. If we attribute this to dust depletion, then we 
can infer that a maximum of 60\% of Fe has gone into dust. In turn, the 
very low abundance of H$_2$ may indicate that the gas is dust free 
with a slight enhancement of $\alpha$ element abundances. 
The chemical properties of this system are very similar to 
the \zabs = 2.402 DLA towards HE~0027$-$1836, that also shows low metallicity and low molecular fraction \citep{Noterdaeme07}.

From Fig.~\ref{highion} it can be seen that C~{\sc iv} and Si~{\sc iv} 
absorption lines are very strong in this DLA. N~{\sc v} is absent 
even though the absorption redshift is close to the QSO emission redshift. 
The absorption lines are seen at the 
expected positions of C~{\sc iii}, N~{\sc iii}, Si~{\sc iii} and
O~{\sc vi} transitions. However, as these transitions are redshifted within the 
Lyman-$\alpha$ forest it is very difficult to confirm them.
This indicates however that a wide range of ionization states
coexist in this system. Absence of  N~{\sc v} and the possible
presence of O~{\sc vi} is consistent with what is typically seen in
DLAs \citep[see][]{Fox07b}.

\subsection{Physical conditions in the absorbing gas}

Here, we discuss the physical conditions in the gas associated 
with system-1 using simple photoionization models using CLOUDY
\citep{Ferland98}.
The aim is not to describe in detail the system but rather to 
estimate the ionization parameter using $N$(Ar~{\sc i}), the particle density 
using $N$(C~{\sc ii$^*$}) and to constrain the distance of the absorbing gas from
the QSO. 

We assume the absorbing cloud to be a plane parallel
slab of constant density, the ionizing radiation to be that
of a typical QSO (i.e \citet{mathews87}), log~$N$(H~{\sc i})~=~21.3 and
Z~=~$-1.4$~Z$_\odot$. The observed dust to metal ratio in the system
is $\sim$ 60\% of the Galactic value. 
Together with the measured metallicity we find the dust to gas ratio 
to be roughly 2\% of what is measured in the cold phase of the neutral 
galactic ISM \citep{welty99a}. We use this value in our model and assume the 
dust composition to be similar to that in the galactic ISM.

\subsubsection{Constraints on the ionization parameter}
\begin{figure}
\centerline{{
\psfig{figure=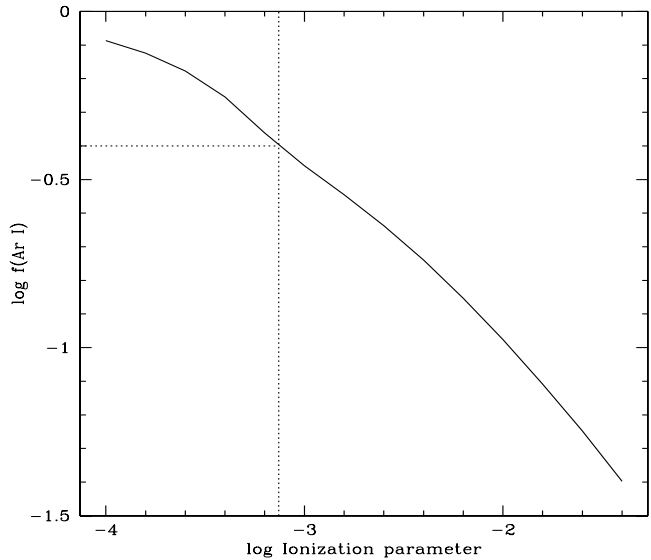,height=8.0cm,width=9.0cm,angle=0}
}}
\caption[]{Fraction of neutral Ar 
as a function of the ionization parameter.
}
\label{fig_ar1}
\end{figure}
\begin{figure}
\centerline{{
\psfig{figure=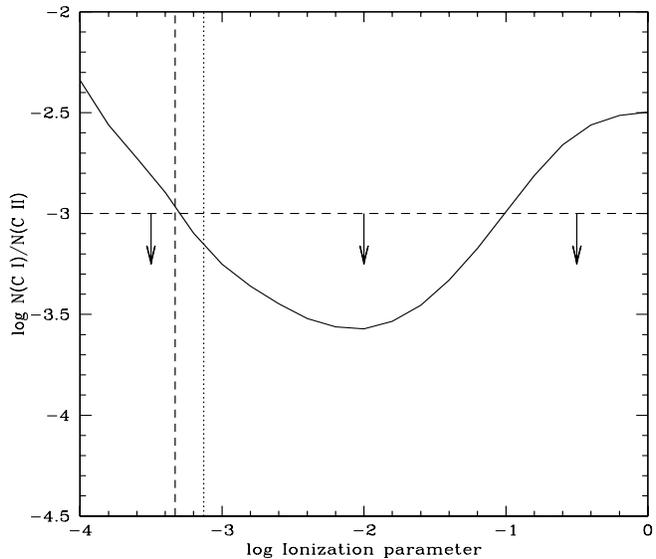,height=8.0cm,width=9.0cm,angle=0}
}}
\caption[]{The predicted $N$(C~{\sc i})/$N$(C~{\sc ii}) ratio as a 
function of ionization parameter. 
}
\label{fig_c1}
\end{figure}

In Fig.~\ref{fig_ar1} we plot the Ar~{\sc i}/Ar ratio as
a function of the ionization parameter, $U$.  
Assuming solar [Ar/S] (and/or [Ar/Si]) abundance ratio, we derive from the
$N$(Ar~{\sc i})/$N$(S~{\sc ii}) (and/or $N$(Ar~{\sc i})/$N$(Si~{\sc ii})) ratio
given in Table~\ref{ave_abundance} that the Ar$^0$/Ar ratio is larger than 0.4.
From Fig.~\ref{fig_ar1} we conclude $U<$~10$^{-3.1}$.
Note that as the average Ar$^0$/Ar ratios 
are similar in different components, this result will hold for 
individual components as well.

In principle we can derive a lower limit on the ionization parameter  
from the upper limit on $N$(C~{\sc i}). As the C~{\sc ii} absorption 
lines available in our spectrum are heavily saturated we have to assume
a value of the [C/Si] abundance ratio. 
From Fig.~3 of \citet{Petitjean08} we note that Si and O abundances follow 
each other very well over a wide range of metallicities. However, 
carbon can be underabundant (by $\sim0.5$ dex) compared to oxygen 
for [O/H]~$\sim$~$-1.4$ as observed in this system.
Using [C/Si]~=~$-0.5$, we derive log~$N$(C~{\sc ii})~=~15.96$\pm$0.15
(i.e log~$N$(Si~{\sc ii})+8.52$-$7.56$-$0.50). This, together
with the 3$\sigma$ upper limit on $N$(C~{\sc i}), gives 
$N$(C~{\sc i})/$N$(C~{\sc ii})~$\le 10^{-3}$.
Fig.~\ref{fig_c1} gives the $N$(C~{\sc i})/$N$(C~{\sc ii}) ratio through the cloud model
versus ionization parameter.
The above condition implies log~$U$~$>$~$-3.3$. 
We therefore find $-3.3\le$log $U\le-3.1$ and we will take
log~$U$~=~$-3.2$ as a typical value for further discussions.

\subsubsection{Fine-structure lines and particle density}
\begin{figure}
\centerline{{
\psfig{figure=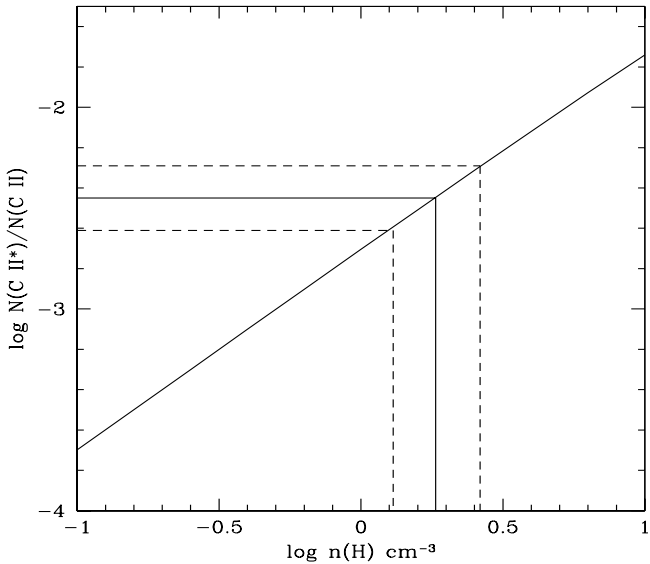,height=8.0cm,width=9.0cm,angle=0}
}}
\caption[]{The slanted solid line gives the 
log $N$(C~{\sc ii}$^*$)/$N$(C~{\sc ii}) ratio as a 
function hydrogen density. The observed mean and 1$\sigma$ ranges for 
$N$(C~{\sc ii}$^*$)/$N$(C~{\sc ii}) are given by horizontal
solid line and dashed lines, respectively. The intersections
of these lines with the model curve give the allowed density range.
}
\label{fig_c2s}
\end{figure}

The only fine-structure absorption line seen in this system is
C~{\sc ii}$^*$$\lambda$1335. Using the same assumptions as above, we can derive 
log~$N$(C~{\sc ii}$^*$)/$N$(C~{\sc ii})~=~$-2.45\pm0.16$ 
from the observations. Interestingly this is consistent
with what is expected in the case of CNM gas with low 
metallicity \citep[See fig. 15 of][]{Srianand05}.

In the absence of C~{\sc i} and associated 
fine-structure lines it is not easy to derive 
$n_{\rm H}$ as the C~{\sc ii$^*$} column density depends on
the ionization fraction of the gas and the kinetic temperature
\citep[see, for example, the discussion in section 6 of][]{Srianand05}.
Therefore, to interpret the $N$(C~{\sc ii$^*$})/$N$(C~{\sc ii}) ratio,
we will rely on a photoionization model.
We run a grid of Cloudy models keeping constant the ionization 
parameter, log~$U$~=~$-3.2$, as constrained above, $T$~=~600 K,
as derived from the 21-cm spin temperature
measurements, and varying the particle density in the range 0.1 to 10~cm$^{-3}$.

The model predictions on $N$(C~{\sc ii}$^*$)/$N$(C~{\sc ii}) 
are given in Fig.~\ref{fig_c2s} as a function of the particle density.  
We derive 1.3~$\le$ $n_{\rm H}$~(cm$^{-3}$)~$\le$~2.6.
Thus a value of 2~cm$^{-3}$ is representative of 
the particle density in this system when we assume log U = $-3.3$. 
Cloudy also predicts the column densities of the 
O~{\sc i} and Si~{\sc ii} fine-structure lines  and 
the predicted values are consistent with the non-detections 
of these species in our
data. 

As an aside, we notice that while most of the constraints 
based on atomic species are satisfied by the model
presented here, the latter predicts high values of $N$(H$_2$) 
(remember we use a dust to gas ratio equal to 
2\% of the Galactic ISM value). However, models
with very low dust content  (i.e dust to gas ratio $\le$0.1\% of 
the Galactic ISM) predict an amount of \h2 consistent with our observations.
Thus it may be possible 
that the observed lower abundance of
Fe co-production elements does not reflect dust depletion but real
nucleosynthetic underabundance. 

\subsubsection{Constraints on the distance to the quasar}

From the estimates of the ionization parameter and particle density, we can
constrain the distance of the absorbing cloud from the QSO. The Lyman limit of the 
absorber in our observed frame is at $\lambda\sim$3807{\AA}. From the
flux calibrated SDSS spectrum we find $f_\nu$~=~70~$\mu$Jy at this
wavelength. For a flat cosmological model with 
$\Omega_\Lambda$~=~0.73, $\Omega_{\rm m}$~=~0.27 and $h = 0.7$ we find
the luminosity distance of the QSO to be 2.8$\times10^{28}$~cm.
This gives the luminosity at the Lyman limit, 
$L_\nu$(LLS)~=~1.7$\times10^{30}$~ergs~s$^{-1}$~Hz$^{-1}$ and the rate of
hydrogen ionizing photons, $Q$(H)~=~2.6$\times 10^{56}$~photons~s$^{-1}$,
when we assume a flat spectrum.
The ionization parameter is usually defined as,
\begin{equation}
U = {\rm Q(H) \over 4 \pi r_{\rm c}^2 c n}.
\end{equation}
Here, $r_{\rm c}$ is the distance of the cloud from the QSO and $n$ is the
particle density in the absorbing gas. Using $n$~$\sim$~2~cm$^{-3}$
and $U$~=~5$\times$10$^{-4}$ we get $r_{\rm c}$~$\sim$~270~kpc.
This also means that, if in situ star formation controls the ionization
state of the gas then the DLA has to be at a distance much larger
than 270~kpc from the QSO.

This clearly shows that the absorbing gas is not associated with the
QSO host galaxy. The gas could be associated with a galaxy in 
the group or cluster in which the QSO resides. In this sense
this system is very similar to 
that of the \zabs = 2.8112 DLA towards PKS~0528$-$250 \citep{Srianand98,Moller98}.

\subsection{Constraining fundamental constants}
As the energy of the hyperfine H~{\sc i} 21-cm transition is 
proportional to the combination of three fundamental constants, 
$x=\alpha^2 G_{\rm p}/ \mu$, 
high resolution optical and 21-cm spectra can be used together to 
probe the combined cosmological variation of these constants \citep{Tubbs80}. 
For the \zabs = 1.7764 system towards Q~1337+170, \citet{Cowie95} have 
obtained $\Delta x/x= (z_{\rm UV}-z_{21})/(1+z_{21}) = (0.7\pm1.1)\times10^{-5}$. Here, $z_{\rm UV}$ and $z_{21}$ are the redshifts
of the UV lines and 21-cm absorption, respectively. 
\citet{tzanavaris05} obtained
${\Delta x/ x} = (0.63\pm0.99)\times 10^{-5}$ 
using this technique and a sample of 8 published systems.
Similarly, \citet{Kanekar06} derived $\Delta x/x$~=~$(-0.15\pm0.62) \times 10^{-5}$
at \zabs = 2.34786 towards PKS B0438$-$438. 

The measured wavelengths of H$_2$ Lyman and Werner band
transitions can be used to probe the variation of $\mu$
\citep{Varshalovich93}. The presence of H$_2$ together with 21-cm absorptions
can, in principle, allow us to place constraints on
$\alpha$ {\sl and} $\mu$ in the same system. In this regard, the system 
studied here is unique.
However, as the \h2 column density is weak, only Werner 
band absorption lines are seen. Therefore and unfortunately, 
the range of sensitivity coefficients of the detected lines 
is too narrow to provide stringent constraints. Using a 
linear regression analysis, we derive $\Delta\mu/\mu \le 4.0 \times 10^{-4}$
and $\bar z_{\rm H_2}$~=~3.174442$\pm$0.000014.
The upper limit on $\Delta\mu/\mu$ measured here is roughly an 
order of magnitude larger than the typical
estimates based on \h2 lines published till now [see \citet{Ivanchik05},\citet{Reinhold06}, \citet{King08} and \citet{Thompson09}].

{As discussed before, the single component fit to the 21-cm absorption
gives $z_{21}$~=~3.174480$\pm$0.000002. 
The strongest metal line component
is seen at $z_{\rm UV}$~=~3.174473$\pm$0.000006. 
From these two redshifts, we derive $\Delta x /x = -(1.7\pm1.5)\times 10^{-6}$. 
We find the wavelength calibration accuracy is $\le 4$ m{\AA} during the
pipeline data reduction
\citep[see also][]{chand06}. This translates to an error of 4$\times 10^{-6}$ 
(or a velocity uncertainty of 0.3 \kms) in
the redshift measurement at the mean observed wavelength ($\sim$ 4500\AA) 
of the lines used to derive $z_{\rm UV}$. If we include 
this error also then 
we get  $\Delta x /x = -(1.7\pm1.7)\times 10^{-6}$.
}

If we assume no variation
in $\alpha$ and $G_{\rm p}$  (respectively $\mu$ and $G_{\rm p}$) then this 
limit 
translates to $\Delta \mu/\mu = -(1.7\pm1.7)\times 10^{-6}$
(respectively $\Delta \alpha/\alpha = -(0.85\pm0.85)\times 10^{-6}$). 
These constraints alone are 
more stringent than most of the published values in the literature.

However, we like to point out that
our UVES spectrum was taken as part of our ongoing survey for molecules
in DLAs where ThAr calibration frames were not taken immediately after
the science observations. This may induce an additional
systematic error of the same order as our measurement errors.
In addition, 
the velocity offset we noticed between \h2 and 21-cm absorption
could reflect inhomogeneities in the gas which could contribute to
additional systematic error.
This kind
of systematics can only be alleviated through randomization using a large and
well defined sample that we want to build in the future.
Note that the velocity shift could be related to covering factor
issues if the radio source shows structures at the milli-arcsec
scale. Thus, as for any other system, high resolution VLBA observations are needed 
to support the analysis towards constraining variations in $x$.

\section{Analysis of system-2 (\zabs~=~3.16768)}

Molecular hydrogen and C~{\sc ii$^*$} absorption
lines are not detected in this DLA with log $N$(H~{\sc i}) = 20.41$\pm$0.15. 
The $2\sigma$ upper limit on
the molecular fraction is $f_{\rm H_2}$~$<$~2$\times 10^{-7}$.
{In addition, as our GMRT observations were performed prior to our UVES observations,
we did not cover the expected frequency 
range for the redshifted 21-cm line}. However, this system is very 
interesting as the metal lines from strong resonance transitions 
are unsaturated and D~{\sc i} absorption can be fitted.
This makes it possible for us to understand the chemical evolution and
deuterium astration factor in this system which is at an early 
stage of chemical evolution.

\subsection{Heavy element abundances and dust}

\begin{figure}
\centerline{{
\psfig{figure=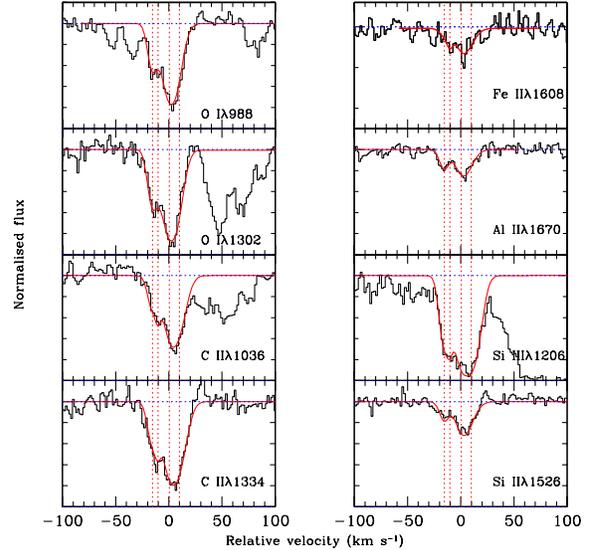,height=8.0cm,width=8.0cm,angle=0}
}}
\caption[]{Multicomponent Voigt profile fits to absorption features
in the \zabs = 3.16768 system. Vertical dotted lines mark the locations of 
individual components. 
}
\label{metal_z2}
\end{figure}

Low ionization metal absorption lines detected in this system
together with the best fitted Voigt profiles 
are shown in Fig.~\ref{metal_z2}. The strongest transitions 
of C~{\sc ii}, Si~{\sc ii} and O~{\sc i} are not strongly saturated and can
be used to derive column densities.
{ The absorption profiles are well fitted with four components at
$v$ = $-$7.7, 0, 12.4 and 17.8 \kms~with respect to \zabs = 3.16768.} 

\begin{table}
\caption{Total column density and average metallicity in 
the \zabs~=~3.16768 DLA}
\begin{center}
\begin{tabular}{lcc}
\hline
\hline
Species & log $N$      & log Z/Z$_\odot^a$ \\
        &  (cm$^{-2}$) &                       \\
\hline
H~{\sc i}     &  20.41$\pm$0.15 & ....\\
N~{\sc i}     & $\le$12.80      &$\le -3.53$ \\
O~{\sc i}     &  14.43$\pm$0.09 &$-2.64\pm0.17$\\
C~{\sc ii}    &  13.98$\pm$0.06 &$-2.82\pm0.16$\\
Si~{\sc ii}   &  13.24$\pm$0.05 &$-2.68\pm0.16$\\
Si~{\sc iii}  &  13.28$\pm$0.06 &\\
Al~{\sc ii}   &  12.00$\pm$0.05 &$-2.78\pm0.16$\\
Fe~{\sc ii}   &  13.14$\pm$0.26 &$-2.72\pm0.30$\\
\hline
\end{tabular}
\begin{flushleft}
{$^a$ solar photospheric abundances are from Asplund et al. (2009)}
\end{flushleft}
\label{tab_met2}
\end{center}
\end{table}

\begin{figure*}
\centerline{{
\psfig{figure=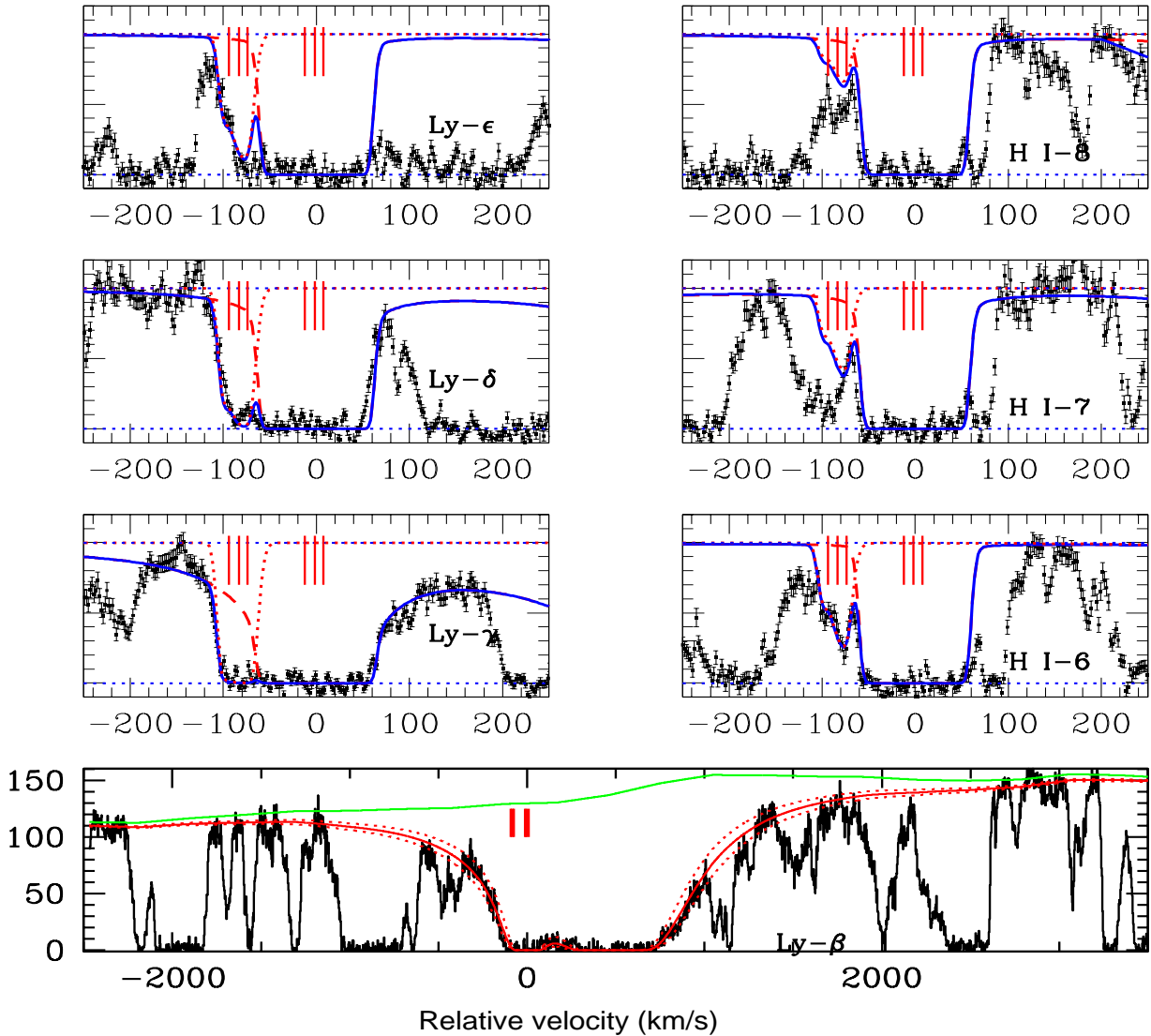,height=16.0cm,width=18.0cm,angle=0}
}}
\caption[]{Voigt profile fits to the Lyman series lines
in the \zabs = 3.16768 system. Ticks around
$v\sim 0$ \kms mark the locations of the metal line components.
The Voigt profile parameters giving good fits
to the damping wings of Lyman-$\beta$, the red wings of Lyman-$\gamma$ and $\delta$ and
the blue wings of Lyman-6, Lyman-7 and Lyman-8 transitions (profiles
with long dashed lines) do not
explain the excess absorption at the expected positions of the
D~{\sc i} absorption lines (ticks around $v = -82$ \kms).
Fits to D~{\sc i} absorption lines are shown with dashed profiles.
}
\label{dbyhfig}
\end{figure*}

Table~\ref{tab_met2} summarises the total column densities of neutral and
low ionization species and  the corresponding metallicities measured in this system
assuming no ionization correction. 
The oxygen metallicity is one of the lowest known among DLAs and sub-DLAs (see
\citet{Petitjean08}; \citet{Pettini08a}). 
We remind the reader that the low oxygen metallicity cannot be  
due to ionization effects because O~{\sc i} is known to be coupled tightly with 
H~{\sc i} through charge exchange reactions. Actually the carbon, silicon, aluminium
and iron metallicities estimated from the singly ionized species are consistent with [O/H].

The system is of low-ionization. We detect the Si~{\sc iii}$\lambda$1206 line 
and measure log~$N$(Si~{\sc iii})~=~13.28$\pm$0.06, but the Si~{\sc iv} absorption 
is very weak (see Fig.~\ref{highion}). Therefore, in any case the silicon 
abundance is [Si/H]~$<$~$-2.38\pm0.13$.

The spectrum covers the expected position of C~{\sc iii}$\lambda$987  
but this line is blended with intervening Lyman-$\alpha$ absorptions (see Fig.~\ref{highion}). 
C~{\sc iv} absorption is also weak. The column density of
C~{\sc ii} implies [C/O]$\ge-0.18\pm0.18$. This value contradicts the result
of standard galactic chemical evolution models 
(see \citet{Akerman04}) expecting the carbon metallicity to be much smaller than
the oxygen metallicity (i.e [C/O]~$<$~$-0.5$) for [O/H]~$<$~$-2$.  
However, our result is consistent with results by 
\cite{Pettini08a} and supports 
the suggestion by \citet{Akerman04} that carbon and oxygen may be produced 
in near-solar proportions in the earliest stages of galactic chemical 
evolution. Chemical evolution models with Population~III stars (\citet{Chieffi02}) or with fast rotating stars (\citet{Chiappini06}) produce near solar
[C/O] at low metallicities.

The spectrum does not show  detectable absorption lines of N~{\sc i}. Hence,
we only derive an upper limit for $N$(N~{\sc i}). The expected wavelength range
for N~{\sc ii}$\lambda$1083 is affected by intervening Lyman-$\alpha$ absorption
from the IGM. Using N~{\sc i}/O~{\sc i} we derive [N/O]$\le$-0.89. This is not
stringent enough to draw conclusions regarding the contribution to the nitrogen
metallicity from primary and secondary productions.

Based on the [Fe/$\alpha$] abundance ratio we can conclude that there
is no dust depletion in the gas. The absence of \h2 absorption 
could very well be attributed to the lack of dust in the system. 
The lack of $\alpha$ element enhancement also means that the star formation 
activity is not recent. In addition, this also undermines the contribution
of Population~III stars to  overall enrichment. 

\subsection{Constraint on the D/H ratio}

While fitting the Lyman series absorption lines, we find a clear
excess absorption in the blue wing of all the available Lyman 
transitions (see Fig.~\ref{dbyhfig}). 
The total H~{\sc i} column density is well
constrained by the damping wings of Lyman-$\beta$ and Lyman-$\gamma$. 
The uncertainty in its measurement comes from the error propagated 
through the Voigt profile fits to the Lyman-$\beta$ absorption of System-1
and from continuum placement. 
{The lower panel of Fig.~\ref{dbyhfig} also shows a typical continuum used
near the wavelength range covered by Lyman-$\beta$ line. This was determined
using the SDSS composite QSO spectrum \citep{VandenBerk01} 
with appropriate scaling to fit the line free regions.} 
We constrain the velocity 
dispersion by using the red wings of Lyman-$\gamma$ and $\delta$ and
using the blue wings of Lyman-6, Lyman-7 and Lyman-8. We notice that
the resulting best fitting Voigt profiles (dotted profiles
in Fig.~\ref{dbyhfig}) under-predict the blue wing 
absorption at $\sim$$-$82 \kms  for Lyman-$\gamma$, $\delta$ and 
$\epsilon$. Consistent absorption dips are present at similar velocity 
in the blue sides of higher Lyman series lines. We do not find
any metal line absorption at this velocity (see Fig.~\ref{metal_z2}).
It is also clear from Fig.~\ref{lyman} that the excess can not be
explained by blending with Lyman series lines from the other two
strong H~{\sc i} absorption systems seen in the immediate vicinity.
Thus it is most likely that the excess absorption in the blue wings
of H~{\sc i} lines
is due to D~{\sc i}.

\begin{figure}
\centerline{\vbox
{
\psfig{figure=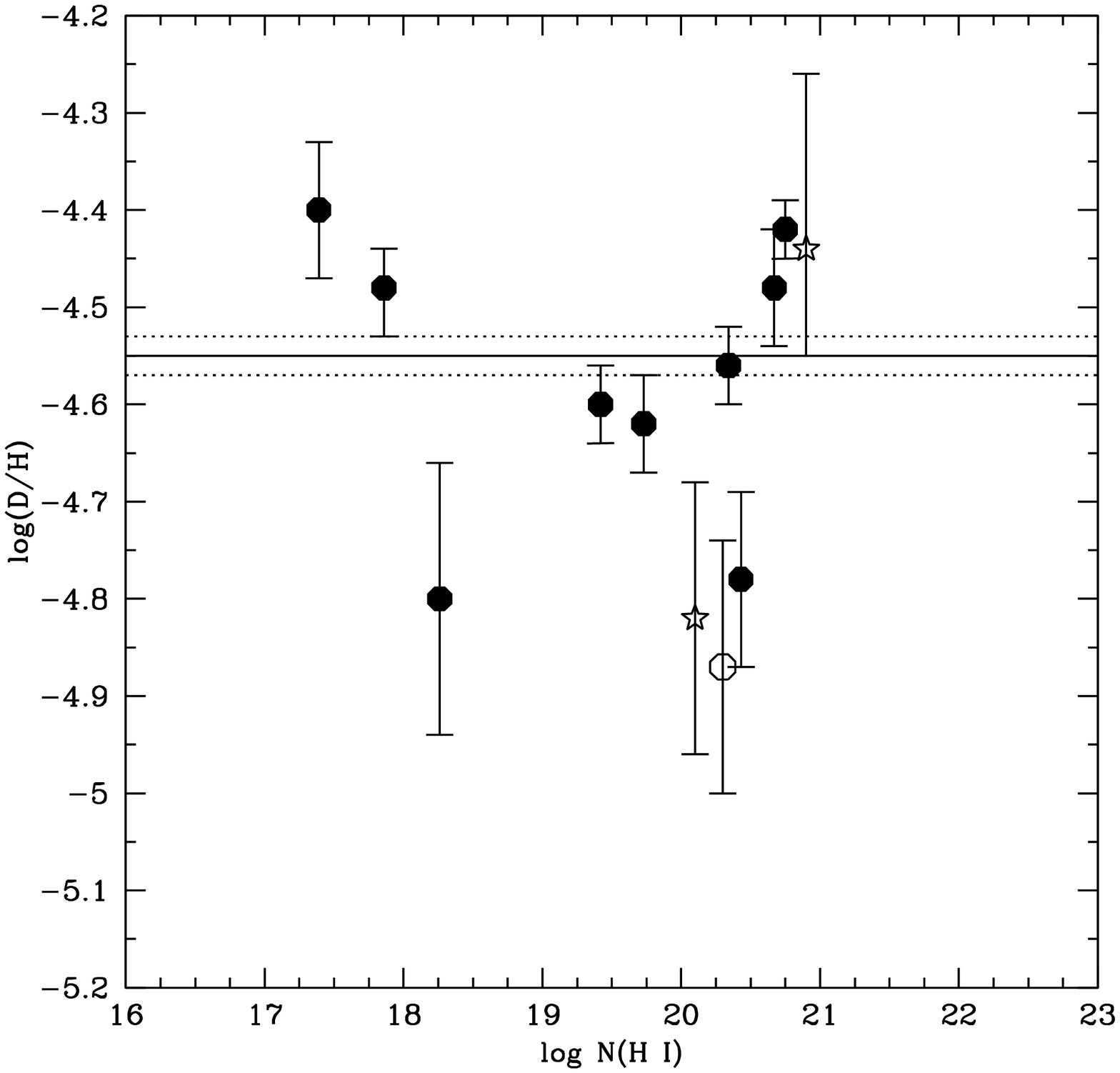,height=7.0cm,width=8.0cm,angle=0}
\psfig{figure=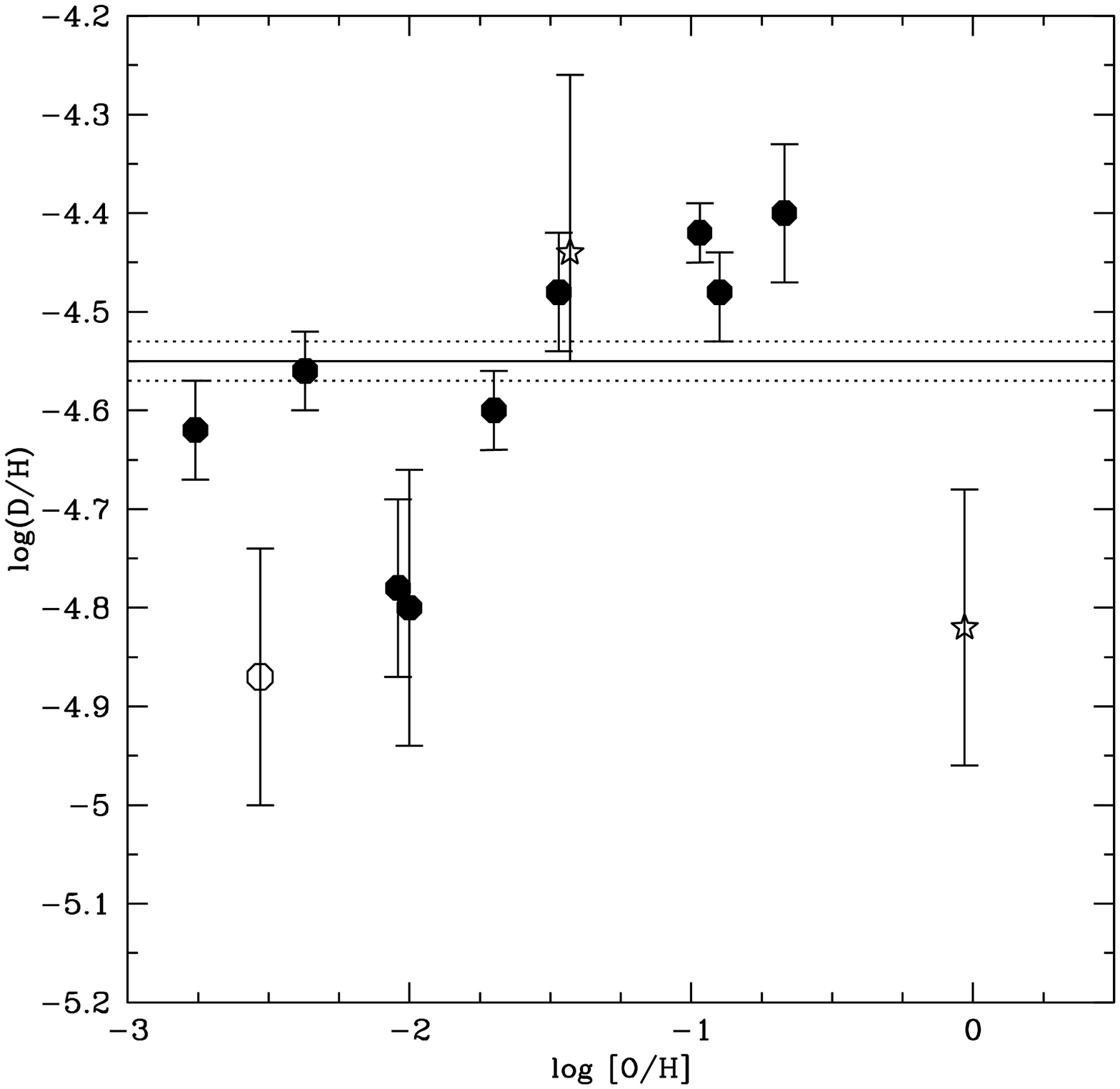,height=7.0cm,width=8.0cm,angle=0}
}}
\caption[]{{\it Top panel:} D/H values measured in absorption systems
as a function of $N$(H~{\sc i}). {\it Bottom panel:} D/H as a function of
metallicity. In both panels the open circle is the measurement presented in
this paper. Filled circles are earlier measurements based on 
$N$(D~{\sc i})/$N$(H~{\sc i}) lines 
from the literature \citep[as summarised in][]{Ivanchik10}. Stars are the 
measurements based on HD/2H$_2$ in two
DLAs presented in \citet{Noterdaeme08hd} and \citet{Ivanchik10}. }
\label{dbyhfig1}
\end{figure}

We simultaneously fit the D~{\sc i} Lyman series lines with three components 
at the redshifts of the three strongest O~{\sc i}
components that contain $\sim 96\%$ of the O~{\sc i} column density.
{We varied $N$ and $b$ for the D~{\sc i} lines to get the best fit.
It is to be noted that D~{\sc i}$-$6 line is unsaturated 
and the total $N$(D~{\sc i}) measurement is insensitive to the exact component
structure.
}
The results of the Voigt profile fits are shown in Fig.~\ref{dbyhfig}.
The best fit model gives $N$(D~{\sc i})~=~$(3.0\pm0.2) \times 10^{15}$~cm$^{-2}$
leading to log~(D/H) = log~(D~{\sc i}/H~{\sc i}) = $-4.93\pm0.15$. This is 
a factor of 2 lower than the primordial value, log~[D/H]$_{\rm p}$ = $-4.59\pm0.02$,
derived from five-year data of the Wilkinson Microwave Anisotropy 
Probe (WMAP)\citep{Komatsu09}. 

For the metallicity estimated in this system, it is believed that [D/H] 
will be close to [D/H]$_{\rm p}$. However, the present system seems to have 
lost an appreciable fraction of deuterium during its evolution. 
In Fig.~\ref{dbyhfig1} we plot D/H measured 
in this system together with those from the literature as a function 
of $N$(H~{\sc i}) and  [O/H]. The literature data includes the
systems listed by \citet{Pettini08b} together with the measurement
by \citet{Crighton04} in a Lyman limit absorption system 
at $z= 3.256$ towards PKS\,1937$-$1009 and the measurement in the \zabs = 3.02 
DLA towards 0347$-$3819 \citep{Levshakov02}. {For this system,
we use the [O/H] measurement given in \citet{Petitjean08}}. We also plot the
D/H measurements based on the HD/H$_2$ ratio by \citet{Noterdaeme08hd} and
\citet{Ivanchik10}. In both panels of Fig.~\ref{dbyhfig1}
the horizontal lines give the weighted mean and 1$\sigma$ range
reported by \citet{Pettini08b} using their sample of 7 measurements.

The latter authors identify two outliers, the \zabs = 2.07623 DLA 
towards Q~2206$-$199 \citep{Pettini01}  and  the \zabs = 2.50357 partial 
Lyman limit system towards Q~1009+299 \citep{Burles98b}, that may be responsible 
for the scatter in their sample. Interestingly, the value measured 
here is consistent with log~(D/H) = $-4.78\pm0.09$ measured in 
the \zabs = 2.07623 metal poor DLA towards Q~2206$-$199.
Therefore, addition of our measurement to the existing sample
increases the scatter (see Fig.~\ref{dbyhfig1}). 
It is also interesting to note that a trend seems to exist for higher 
D/H values at higher oxygen metallicity which may be difficult to understand. 

As the absorption lines of H~{\sc i} and D~{\sc i} 
fall inside the Lyman-$\alpha$ 
forest one can always question
the accuracy of column density measurements.
As we use damping wings, the $N$(H~{\sc i}) measurement is robust 
as is our lower limit on the astration factor. Indeed, if the deuterium absorption
features are due to intervening Lyman-$\alpha$ clouds, the astration
factor must be even larger.

There are recent indications that infalling gas may cause
astration factor to be different from that predicted
by chemical evolution closed box models.                       
\citet{Noterdaeme08hd} measured a relatively low astration factor 
of deuterium coupled
with a high metal enrichment in the  \zabs = 2.418 DLA 
towards Q~1439+1117. 
\citet{Ivanchik10} have reported a [D/H] value close to [D/H]$_{\rm p}$
for the  \zabs = 2.338 DLA towards Q~1232+0815 based on \h2 and HD
column densities.
This system has [O/H] = $-$1.43$\pm$0.08 \citep{Srianand00}. 
Thus it appears that the astration factor of deuterium varies largely from one system to 
the other at a given metallicity.
Therefore, it may not be a good idea to remove the so-called outliers while computing 
the weighted mean value of [D/H] measured in absorption line systems.

\section{Conclusion}
We have presented detailed analysis of two very closely spaced DLAs 
(at \zabs = 3.17447 and 3.16768) within a velocity separation of 
$\sim$500 \kms along the line of sight to SDSS~J1337+3152.
We have reported the detection of 21-cm and molecular hydrogen absorption 
lines in the damped Lyman-$\alpha$ system at \zabs = 3.17447. 
This is the second time that both species are detected in the 
same system. This is a unique combination to constrain at the same time
the time variation of the fine structure constant, $\alpha$, and the
proton-to-electron mass ratio, $\mu$. We note however
a velocity shift of $\sim$2.7~km/s between the \h2 and 21-cm 
absorption, revealing 
small scale inhomogeneities in the cloud that could wash out detection of
a variation in $x$. Such systematics can only be
alleviated with a large sample of such systems. However, the redshift
of the strongest metal line component matches well with the redshifted
21-cm absorption line. Using this, we derive  a stringent limit on the variation of
$x=\alpha^2 g_{\rm p}/ \mu$, $\Delta x /x = -(1.7\pm1.7)\times 10^{-6}$, 
which is one of the best limits ever obtained. While this demonstrates
the advantages of combining GMRT and UVES data,
we emphasize the need for a large sample of such measurements.

Using $N$(H~{\sc i}) measured from Voigt profile fitting of the damped Lyman-$\alpha$
line in the UVES spectrum together with the 21-cm optical depth measured from our 
GMRT spectrum, we obtain
$T_{\rm S}/f_{\rm c} = 600^{+222}_{-159}$ K. This, together with the velocity shift
noted above, is consistent with the absorbing gas being a mixture of different phases.
We derive mean physical conditions in the cloud, $U$~$\sim$~10$^{-3.2}$, 
$n_{\rm H}$~$\sim$~2~cm$^{-3}$, $T_{\rm K}$~$\sim$~600~K. The metallicity is
of the order of [S/H]~$\sim$~$-1.5$ and the distance to the quasar larger
than 270~kpc. We conclude that the gas must be associated with a galaxy
in the group or cluster in which the QSO resides.

The gas responsible for the \zabs = 3.16768 DLA has very low metallicity,
[O/H]~=~$-2.64\pm0.17$. 
This allows us to measure the column density of C~{\sc ii} accurately.
{ We measure [C/O] = -0.18$\pm$0.18}. This
means that 
[C/O] reaches solar value even at very low metallicities.
We measure the deuterium abundance in this DLA to be 
log~(D/H) = log~(D~{\sc i}/H~{\sc i}) = $-4.93\pm0.15$. This is 
a factor of 2 lower than the primordial value, [D/H]$_{\rm p}$ = $-4.59\pm0.02$,
derived from five-year data of the Wilkinson Microwave Anisotropy 
Probe (WMAP) \citep{Komatsu09}. 
We therefore conclude that astration factors can vary significantly even at low metallicity.

\section*{acknowledgements}
We thank the GMRT staff for their co-operation during our observations.
The GMRT is an international facility run by the National Centre for Radio 
Astrophysics of the Tata Institute of Fundamental Research.
We acknowledge the use of SDSS spectra from the archive (http://www.sdss.org/).
PN acknowledges the support from the french Ministry of Foreign and European Affairs.

\def\aj{AJ}%
\def\actaa{Acta Astron.}%
\def\araa{ARA\&A}%
\def\apj{ApJ}%
\def\apjl{ApJ}%
\def\apjs{ApJS}%
\def\ao{Appl.~Opt.}%
\def\apss{Ap\&SS}%
\def\aap{A\&A}%
\def\aapr{A\&A~Rev.}%
\def\aaps{A\&AS}%
\def\azh{AZh}%
\def\baas{BAAS}%
\def\bac{Bull. astr. Inst. Czechosl.}%
\def\caa{Chinese Astron. Astrophys.}%
\def\cjaa{Chinese J. Astron. Astrophys.}%
\def\icarus{Icarus}%
\def\jcap{J. Cosmology Astropart. Phys.}%
\def\jrasc{JRASC}%
\def\mnras{MNRAS}%
\def\memras{MmRAS}%
\def\na{New A}%
\def\nar{New A Rev.}%
\def\pasa{PASA}%
\def\pra{Phys.~Rev.~A}%
\def\prb{Phys.~Rev.~B}%
\def\prc{Phys.~Rev.~C}%
\def\prd{Phys.~Rev.~D}%
\def\pre{Phys.~Rev.~E}%
\def\prl{Phys.~Rev.~Lett.}%
\def\pasp{PASP}%
\def\pasj{PASJ}%
\def\qjras{QJRAS}%
\def\rmxaa{Rev. Mexicana Astron. Astrofis.}%
\def\skytel{S\&T}%
\def\solphys{Sol.~Phys.}%
\def\sovast{Soviet~Ast.}%
\def\ssr{Space~Sci.~Rev.}%
\def\zap{ZAp}%
\def\nat{Nature}%
\def\iaucirc{IAU~Circ.}%
\def\aplett{Astrophys.~Lett.}%
\def\apspr{Astrophys.~Space~Phys.~Res.}%
\def\bain{Bull.~Astron.~Inst.~Netherlands}%
\def\fcp{Fund.~Cosmic~Phys.}%
\def\gca{Geochim.~Cosmochim.~Acta}%
\def\grl{Geophys.~Res.~Lett.}%
\def\jcp{J.~Chem.~Phys.}%
\def\jgr{J.~Geophys.~Res.}%
\def\jqsrt{J.~Quant.~Spec.~Radiat.~Transf.}%
\def\memsai{Mem.~Soc.~Astron.~Italiana}%
\def\nphysa{Nucl.~Phys.~A}%
\def\physrep{Phys.~Rep.}%
\def\physscr{Phys.~Scr}%
\def\planss{Planet.~Space~Sci.}%
\def\procspie{Proc.~SPIE}%
\let\astap=\aap
\let\apjlett=\apjl
\let\apjsupp=\apjs
\let\applopt=\ao
\bibliographystyle{mn}

\begin{thebibliography}{61}
\expandafter\ifx\csname natexlab\endcsname\relax\def\natexlab#1{#1}\fi

\bibitem[{{Abgrall} {et~al.}(1994){Abgrall}, {Roueff}, {Launay}, \&
  {Roncin}}]{Abgrall94}
{Abgrall}, H., {Roueff}, E., {Launay}, F., \& {Roncin}, J.-Y., 1994, Canadian
  Journal of Physics, 72, 856

\bibitem[{{Akerman} {et~al.}(2004){Akerman}, {Carigi}, {Nissen}, {Pettini}, \&
  {Asplund}}]{Akerman04}
{Akerman}, C.~J., {Carigi}, L., {Nissen}, P.~E., {Pettini}, M., \& {Asplund},
  M., 2004, \aap, 414, 931

\bibitem[{{Asplund} {et~al.}(2009){Asplund}, {Grevesse}, {Sauval}, \&
  {Scott}}]{Asplund09}
{Asplund}, M., {Grevesse}, N., {Sauval}, A.~J., \& {Scott}, P., 2009, \araa,
  47, 481

\bibitem[{{Ballester} {et~al.}(2000){Ballester}, {Modigliani}, {Boitquin},
  {Cristiani}, {Hanuschik}, {Kauffer}, \& {Wolf}}]{Ballester00}
{Ballester}, P., {Modigliani}, A., {Boitquin}, O., {Cristiani}, S.,
  {Hanuschik}, R., {Kauffer}, A., \& {Wolf}, S., 2000, The Messenger, 101, 31

\bibitem[{{Burles} \& {Tytler}(1998)}]{Burles98b}
{Burles}, S. \& {Tytler}, D., 1998, \apj, 507, 732

\bibitem[{{Carilli} \& {van Gorkom}(1992)}]{Carilli92}
{Carilli}, C.~L. \& {van Gorkom}, J.~H., 1992, \apj, 399, 373

\bibitem[{{Chand} {et~al.}(2006){Chand}, {Srianand}, {Petitjean}, {Aracil},
  {Quast}, \& {Reimers}}]{chand06}
{Chand}, H., {Srianand}, R., {Petitjean}, P., {Aracil}, B., {Quast}, R., \&
  {Reimers}, D., 2006, \aap, 451, 45

\bibitem[{{Chiappini} {et~al.}(2006){Chiappini}, {Hirschi}, {Meynet},
  {Ekstr{\"o}m}, {Maeder}, \& {Matteucci}}]{Chiappini06}
{Chiappini}, C., {Hirschi}, R., {Meynet}, G., {Ekstr{\"o}m}, S., {Maeder}, A.,
  \& {Matteucci}, F., 2006, \aap, 449, L27

\bibitem[{{Chieffi} \& {Limongi}(2002)}]{Chieffi02}
{Chieffi}, A. \& {Limongi}, M., 2002, \apj, 577, 281

\bibitem[{{Cowie} \& {Songaila}(1995)}]{Cowie95}
{Cowie}, L.~L. \& {Songaila}, A., 1995, \apj, 453, 596

\bibitem[{{Crighton} {et~al.}(2004){Crighton}, {Webb}, {Ortiz-Gil}, \&
  {Fern{\'a}ndez-Soto}}]{Crighton04}
{Crighton}, N.~H.~M., {Webb}, J.~K., {Ortiz-Gil}, A., \& {Fern{\'a}ndez-Soto},
  A., 2004, \mnras, 355, 1042

\bibitem[{{Cui} {et~al.}(2005){Cui}, {Bechtold}, {Ge}, \& {Meyer}}]{Cui05}
{Cui}, J., {Bechtold}, J., {Ge}, J., \& {Meyer}, D.~M., 2005, \apj, 633, 649

\bibitem[{{Dekker} {et~al.}(2000){Dekker}, {D'Odorico}, {Kaufer}, {Delabre}, \&
  {Kotzlowski}}]{Dekker00}
{Dekker}, H., {D'Odorico}, S., {Kaufer}, A., {Delabre}, B., \& {Kotzlowski},
  H., 2000, in Proc. SPIE Vol. 4008, p. 534-545, Optical and IR Telescope
  Instrumentation and Detectors, Masanori Iye; Alan F. Moorwood; Eds., pp.
  534--545

\bibitem[{{Ferland} {et~al.}(1998){Ferland}, {Korista}, {Verner}, {Ferguson},
  {Kingdon}, \& {Verner}}]{Ferland98}
{Ferland}, G.~J., {Korista}, K.~T., {Verner}, D.~A., {Ferguson}, J.~W.,
  {Kingdon}, J.~B., \& {Verner}, E.~M., 1998, \pasp, 110, 761

\bibitem[{{Fox} {et~al.}(2007){Fox}, {Petitjean}, {Ledoux}, \&
  {Srianand}}]{Fox07b}
{Fox}, A.~J., {Petitjean}, P., {Ledoux}, C., \& {Srianand}, R., 2007, \aap,
  465, 171

\bibitem[{{Gupta} {et~al.}(2006){Gupta}, {Salter}, {Saikia}, {Ghosh}, \&
  {Jeyakumar}}]{Gupta06}
{Gupta}, N., {Salter}, C.~J., {Saikia}, D.~J., {Ghosh}, T., \& {Jeyakumar}, S.,
  2006, \mnras, 373, 972

\bibitem[{{Gupta} {et~al.}(2009){Gupta}, {Srianand}, {Petitjean}, {Noterdaeme},
  \& {Saikia}}]{gupta09}
{Gupta}, N., {Srianand}, R., {Petitjean}, P., {Noterdaeme}, P., \& {Saikia},
  D.~J., 2009, \mnras, 398, 201

\bibitem[{{Hamann} \& {Ferland}(1993)}]{Hamann93}
{Hamann}, F. \& {Ferland}, G., 1993, \apj, 418, 11

\bibitem[{{Heinm{\"u}ller} {et~al.}(2006){Heinm{\"u}ller}, {Petitjean},
  {Ledoux}, {Caucci}, \& {Srianand}}]{Heinmuller06}
{Heinm{\"u}ller}, J., {Petitjean}, P., {Ledoux}, C., {Caucci}, S., \&
  {Srianand}, R., 2006, \aap, 449, 33

\bibitem[{{Ivanchik} {et~al.}(2005){Ivanchik}, {Petitjean}, {Varshalovich},
  {Aracil}, {Srianand}, {Chand}, {Ledoux}, \& {Boiss{\'e}}}]{Ivanchik05}
{Ivanchik}, A., {Petitjean}, P., {Varshalovich}, D., {Aracil}, B., {Srianand},
  R., {Chand}, H., {Ledoux}, C., \& {Boiss{\'e}}, P., 2005, \aap, 440, 45

\bibitem[{{Ivanchik} {et~al.}(2010){Ivanchik}, {Petitjean}, {Balashev},
  {Srianand}, {Varshalovich}, {Ledoux}, \& {Noterdaeme}}]{Ivanchik10}
{Ivanchik}, A.~V., {Petitjean}, P., {Balashev}, S.~A., {Srianand}, R.,
  {Varshalovich}, D.~A., {Ledoux}, C., \& {Noterdaeme}, P., 2010, ArXiv
  e-prints

\bibitem[{{Kanekar} \& {Chengalur}(2003)}]{kanekar03}
{Kanekar}, N. \& {Chengalur}, J.~N., 2003, \aap, 399, 857

\bibitem[{{Kanekar} {et~al.}(2006){Kanekar}, {Subrahmanyan}, {Ellison}, {Lane},
  \& {Chengalur}}]{Kanekar06}
{Kanekar}, N., {Subrahmanyan}, R., {Ellison}, S.~L., {Lane}, W.~M., \&
  {Chengalur}, J.~N., 2006, \mnras, 370, L46

\bibitem[{{King} {et~al.}(2008){King}, {Webb}, {Murphy}, \&
  {Carswell}}]{King08}
{King}, J.~A., {Webb}, J.~K., {Murphy}, M.~T., \& {Carswell}, R.~F., 2008,
  Physical Review Letters, 101, 251304

\bibitem[{{Komatsu} {et~al.}(2009){Komatsu}, {Dunkley}, {Nolta}, {Bennett},
  {Gold}, {Hinshaw}, {Jarosik}, {Larson}, {Limon}, {Page}, {Spergel},
  {Halpern}, {Hill}, {Kogut}, {Meyer}, {Tucker}, {Weiland}, {Wollack}, \&
  {Wright}}]{Komatsu09}
{Komatsu}, E., {Dunkley}, J., {Nolta}, M.~R., {et~al.}, 2009, \apjs, 180, 330

\bibitem[{{Ledoux} {et~al.}(2003){Ledoux}, {Petitjean}, \&
  {Srianand}}]{ledoux03}
{Ledoux}, C., {Petitjean}, P., \& {Srianand}, R., 2003, \mnras, 346, 209

\bibitem[{{Ledoux} {et~al.}(2006){Ledoux}, {Petitjean}, \&
  {Srianand}}]{Ledoux06b}
---, 2006, \apjl, 640, L25

\bibitem[{{Levshakov} {et~al.}(2002){Levshakov}, {Dessauges-Zavadsky},
  {D'Odorico}, \& {Molaro}}]{Levshakov02}
{Levshakov}, S.~A., {Dessauges-Zavadsky}, M., {D'Odorico}, S., \& {Molaro}, P.,
  2002, \apj, 565, 696

\bibitem[{{Markwardt}(2009)}]{Markwardt09}
{Markwardt}, C.~B., 2009, ArXiv e-prints 0902.2850

\bibitem[{{Mathews} \& {Ferland}(1987)}]{mathews87}
{Mathews}, W.~G. \& {Ferland}, G.~J., 1987, \apj, 323, 456

\bibitem[{{Moller} \& {Warren}(1998)}]{Moller98}
{Moller}, P. \& {Warren}, S.~J., 1998, \mnras, 299, 661

\bibitem[{{Morton}(2003)}]{Morton03}
{Morton}, D.~C., 2003, \apjs, 149, 205

\bibitem[{{Noterdaeme} {et~al.}(2008{\natexlab{a}}){Noterdaeme}, {Ledoux},
  {Petitjean}, \& {Srianand}}]{noterdaeme08}
{Noterdaeme}, P., {Ledoux}, C., {Petitjean}, P., \& {Srianand}, R.,
  2008{\natexlab{a}}, \aap, 481, 327

\bibitem[{{Noterdaeme} {et~al.}(2009{\natexlab{a}}){Noterdaeme}, {Ledoux},
  {Srianand}, {Petitjean}, \& {Lopez}}]{noterdaeme09co}
{Noterdaeme}, P., {Ledoux}, C., {Srianand}, R., {Petitjean}, P., \& {Lopez},
  S., 2009{\natexlab{a}}, \aap, 503, 765

\bibitem[{{Noterdaeme} {et~al.}(2009{\natexlab{b}}){Noterdaeme}, {Petitjean},
  {Ledoux}, \& {Srianand}}]{Noterdaeme09dla}
{Noterdaeme}, P., {Petitjean}, P., {Ledoux}, C., \& {Srianand}, R.,
  2009{\natexlab{b}}, \aap, 505, 1087

\bibitem[{{Noterdaeme} {et~al.}(2008{\natexlab{b}}){Noterdaeme}, {Petitjean},
  {Ledoux}, {Srianand}, \& {Ivanchik}}]{Noterdaeme08hd}
{Noterdaeme}, P., {Petitjean}, P., {Ledoux}, C., {Srianand}, R., \& {Ivanchik},
  A., 2008{\natexlab{b}}, \aap, 491, 397

\bibitem[{{Noterdaeme} {et~al.}(2007){Noterdaeme}, {Petitjean}, {Srianand},
  {Ledoux}, \& {Le Petit}}]{Noterdaeme07}
{Noterdaeme}, P., {Petitjean}, P., {Srianand}, R., {Ledoux}, C., \& {Le Petit},
  F., 2007, \aap, 469, 425

\bibitem[{{Petitjean} {et~al.}(2008){Petitjean}, {Ledoux}, \&
  {Srianand}}]{Petitjean08}
{Petitjean}, P., {Ledoux}, C., \& {Srianand}, R., 2008, \aap, 480, 349

\bibitem[{{Petitjean} {et~al.}(1994){Petitjean}, {Rauch}, \&
  {Carswell}}]{Petitjean94}
{Petitjean}, P., {Rauch}, M., \& {Carswell}, R.~F., 1994, \aap, 291, 29

\bibitem[{{Petitjean} {et~al.}(2000){Petitjean}, {Srianand}, \&
  {Ledoux}}]{petitjean00}
{Petitjean}, P., {Srianand}, R., \& {Ledoux}, C., 2000, \aap, 364, L26

\bibitem[{{Pettini} \& {Bowen}(2001)}]{Pettini01}
{Pettini}, M. \& {Bowen}, D.~V., 2001, \apj, 560, 41

\bibitem[{{Pettini} {et~al.}(2008{\natexlab{a}}){Pettini}, {Zych}, {Murphy},
  {Lewis}, \& {Steidel}}]{Pettini08b}
{Pettini}, M., {Zych}, B.~J., {Murphy}, M.~T., {Lewis}, A., \& {Steidel},
  C.~C., 2008{\natexlab{a}}, \mnras, 391, 1499

\bibitem[{{Pettini} {et~al.}(2008{\natexlab{b}}){Pettini}, {Zych}, {Steidel},
  \& {Chaffee}}]{Pettini08a}
{Pettini}, M., {Zych}, B.~J., {Steidel}, C.~C., \& {Chaffee}, F.~H.,
  2008{\natexlab{b}}, \mnras, 385, 2011

\bibitem[{{Reinhold} {et~al.}(2006){Reinhold}, {Buning}, {Hollenstein},
  {Ivanchik}, {Petitjean}, \& {Ubachs}}]{Reinhold06}
{Reinhold}, E., {Buning}, R., {Hollenstein}, U., {Ivanchik}, A., {Petitjean},
  P., \& {Ubachs}, W., 2006, \prl, 96, 151101

\bibitem[{{Rodr{\'{\i}}guez} {et~al.}(2006){Rodr{\'{\i}}guez}, {Petitjean},
  {Aracil}, {Ledoux}, \& {Srianand}}]{Rodriguez06}
{Rodr{\'{\i}}guez}, E., {Petitjean}, P., {Aracil}, B., {Ledoux}, C., \&
  {Srianand}, R., 2006, \aap, 446, 791

\bibitem[{{Roy} {et~al.}(2006){Roy}, {Chengalur}, \& {Srianand}}]{Roy06}
{Roy}, N., {Chengalur}, J.~N., \& {Srianand}, R., 2006, \mnras, 365, L1

\bibitem[{{Sofia} \& {Jenkins}(1998)}]{Sofia98}
{Sofia}, U.~J. \& {Jenkins}, E.~B., 1998, \apj, 499, 951

\bibitem[{{Srianand} {et~al.}(2008){Srianand}, {Noterdaeme}, {Ledoux}, \&
  {Petitjean}}]{srianand08}
{Srianand}, R., {Noterdaeme}, P., {Ledoux}, C., \& {Petitjean}, P., 2008, \aap,
  482, L39

\bibitem[{{Srianand} \& {Petitjean}(1998)}]{Srianand98}
{Srianand}, R. \& {Petitjean}, P., 1998, \aap, 335, 33

\bibitem[{{Srianand} {et~al.}(2000){Srianand}, {Petitjean}, \&
  {Ledoux}}]{Srianand00}
{Srianand}, R., {Petitjean}, P., \& {Ledoux}, C., 2000, \nat, 408, 931

\bibitem[{{Srianand} {et~al.}(2005){Srianand}, {Petitjean}, {Ledoux},
  {Ferland}, \& {Shaw}}]{Srianand05}
{Srianand}, R., {Petitjean}, P., {Ledoux}, C., {Ferland}, G., \& {Shaw}, G.,
  2005, \mnras, 362, 549

\bibitem[{{Thompson} {et~al.}(2009){Thompson}, {Bechtold}, {Black},
  {Eisenstein}, {Fan}, {Kennicutt}, {Martins}, {Prochaska}, \&
  {Shirley}}]{Thompson09}
{Thompson}, R.~I., {Bechtold}, J., {Black}, J.~H., {et~al.}, 2009, \apj, 703,
  1648

\bibitem[{{Tubbs} \& {Wolfe}(1980)}]{Tubbs80}
{Tubbs}, A.~D. \& {Wolfe}, A.~M., 1980, \apjl, 236, L105

\bibitem[{{Tzanavaris} {et~al.}(2005){Tzanavaris}, {Webb}, {Murphy},
  {Flambaum}, \& {Curran}}]{tzanavaris05}
{Tzanavaris}, P., {Webb}, J.~K., {Murphy}, M.~T., {Flambaum}, V.~V., \&
  {Curran}, S.~J., 2005, Physical Review Letters, 95, 041301

\bibitem[{{Ubachs} {et~al.}(2007){Ubachs}, {Buning}, {Eikema}, \&
  {Reinhold}}]{Ubachs07}
{Ubachs}, W., {Buning}, R., {Eikema}, K.~S.~E., \& {Reinhold}, E., 2007,
  Journal of Molecular Spectroscopy, 241, 155

\bibitem[{{Vanden Berk} {et~al.}(2001){Vanden Berk}, {Richards}, {Bauer},
  {Strauss}, {Schneider}, {Heckman}, {York}, {Hall}, {Fan}, {Knapp},
  {Anderson}, {Annis}, {Bahcall}, {Bernardi}, {Briggs}, {Brinkmann}, {Brunner},
  {Burles}, {Carey}, {Castander}, {Connolly}, {Crocker}, {Csabai}, {Doi},
  {Finkbeiner}, {Friedman}, {Frieman}, {Fukugita}, {Gunn}, {Hennessy},
  {Ivezi{\'c}}, {Kent}, {Kunszt}, {Lamb}, {Leger}, {Long}, {Loveday}, {Lupton},
  {Meiksin}, {Merelli}, {Munn}, {Newberg}, {Newcomb}, {Nichol}, {Owen}, {Pier},
  {Pope}, {Rockosi}, {Schlegel}, {Siegmund}, {Smee}, {Snir}, {Stoughton},
  {Stubbs}, {SubbaRao}, {Szalay}, {Szokoly}, {Tremonti}, {Uomoto}, {Waddell},
  {Yanny}, \& {Zheng}}]{VandenBerk01}
{Vanden Berk}, D.~E., {Richards}, G.~T., {Bauer}, A., {et~al.}, 2001, \aj, 122,
  549

\bibitem[{{Varshalovich} \& {Levshakov}(1993)}]{Varshalovich93}
{Varshalovich}, D.~A. \& {Levshakov}, S.~A., 1993, Soviet Journal of
  Experimental and Theoretical Physics Letters, 58, 237

\bibitem[{{Vladilo} {et~al.}(2003){Vladilo}, {Centuri{\'o}n}, {D'Odorico}, \&
  {P{\'e}roux}}]{Vladilo03}
{Vladilo}, G., {Centuri{\'o}n}, M., {D'Odorico}, V., \& {P{\'e}roux}, C., 2003,
  \aap, 402, 487

\bibitem[{{Welty} {et~al.}(1999){Welty}, {Hobbs}, {Lauroesch}, {Morton},
  {Spitzer}, \& {York}}]{welty99a}
{Welty}, D.~E., {Hobbs}, L.~M., {Lauroesch}, J.~T., {Morton}, D.~C., {Spitzer},
  L., \& {York}, D.~G., 1999, \apjs, 124, 465

\bibitem[{{Wolfe} {et~al.}(2005){Wolfe}, {Gawiser}, \& {Prochaska}}]{wolfe05}
{Wolfe}, A.~M., {Gawiser}, E., \& {Prochaska}, J.~X., 2005, \araa, 43, 861

\bibitem[{{Wolfire} {et~al.}(2003){Wolfire}, {McKee}, {Hollenbach}, \&
  {Tielens}}]{Wolfire03}
{Wolfire}, M.~G., {McKee}, C.~F., {Hollenbach}, D., \& {Tielens}, A.~G.~G.~M.,
  2003, \apj, 587, 278

\end{thebibliography}

\end{document}